\documentclass[aps,prb,showpacs,twocolumn]{revtex4-1}

\usepackage{natbib}
\usepackage{amsmath}
\usepackage{amssymb}
\usepackage{bm}
\usepackage{graphicx}
\usepackage{times}
\usepackage{wrapfig}
\usepackage{array}
\usepackage{color}
\usepackage{arydshln}
\usepackage{accents}

\graphicspath{{plots/},{figures/}}

\begin{document}
\title{Higher-order contributions and non-perturbative effects in the non-degenerate nonlinear optical absorption of direct-gap semiconductors}
\author{W.-R.~Hannes and T.~Meier}
\affiliation{Department of Physics and CeOPP, University of Paderborn, Warburger Str.~100, D-33098 Paderborn, Germany}

\date{\today}

\begin{abstract}
The semiconductor Bloch equations for a two-band model including inter- and intraband excitation are used to study the nonlinear absorption of single and multiple light pulses by direct-gap semiconductors.
For a consistent analysis the contributions to the absorption originating from both the interband polarization and the intraband current need to be included.
In the Bloch equation approach theses contributions as well as different excitation pathways in terms of sequences of inter- and intraband excitations can be evaluated separately which allows for a transparent analysis, the identification of the dominant terms, and analyzing their dependence on the excitation conditions.
In the perturbative regime, we obtain analytical expressions for the multi-photon absorption coefficients for continuous-wave excitation. These results are shown to agree well with numerical results for short pulses and/or finite dephasing and relaxation times and we confirm the previously predicted strong enhancement of two-photon absorption for non-degenerate conditions for pulsed excitation. We discuss the dependencies on the light frequencies, initial band populations, and the time delay between the pulses.
The frequency dependence of the two-photon absorption coefficient for non-degenerate excitation is evaluated perturbatively in third-order. The higher-order contributions to the optical absorption include three- and four-photon absorption and show a rich frequency dependence including negative regions and dispersive lineshapes.
Non-perturbative solutions of the Bloch equations demonstrate a strongly non-monotonous behavior of the intensity-dependent optical absorption for a single incident pulse and in a pump-probe set-up.
\end{abstract}


\maketitle	

\section{Introduction}

Two- and multi-photon absorption (2PA/MPA) are among the most fundamental effects
\cite{Goeppert‐Mayer1931,Braunstein1964,Loudon1962,Hopfield1965,Mahan1968} in nonlinear optics that has been proposed and studied for several decades considering various physical systems.
Applications include infrared detection,\cite{Schneider2005,Schneider2009,Fishman2011,Pattanaik2016a,Piccardo2018}
optical switching,\cite{Venkataraman2011,Yavuz2006}
and optical control of electrical currents.\cite{Atanasov1996}
The time-reversed process of stimulated two-photon emission\cite{Hayat2008,Reichert2016} is expected to eventually lead to the realization of two-photon lasing.
For solid-state systems this topic is currently studied intensively, also with regards to applications in the ultrafast and strongly nonlinear regimes.
However, presently still several fundamental aspect for MPA in solids are not well understood, in particular, with regards to higher-order processes and the dynamical behavior.

Traditionally 2PA has been studied mainly by transition rate approaches.\cite{Braunstein1964,Hopfield1965,Loudon1962,Basov1966,Mahan1968}
Even simplistic band structure models lead to reasonable predictions of 2PA spectra for various semiconductors.\cite{Bolger1993,Said1992,Sheik-Bahae1991}
However, transition rate approaches are not easy to evaluate for higher numbers of simultaneously absorbed photons and also do not account for dephasing or relaxation effects. 
A more versatile approach describing the nonlinear-optical response of semiconductors is based on susceptibilities. 
Typically this approach is applied in a semiclassical and/or an adiabatic approximation. 
In Refs.~\onlinecite{Sipe1993,Aversa1994,Aversa1995} it has been evaluated in the framework of a semiclassical density matrix formalism for the steady-state nonlinear-optical response.

In this study we use a more general approach based on the semiconductor Bloch equations (SBE) extended by the intraband terms,\cite{Haug2009,Meier1994} thereby neglecting electron-electron interactions. 
Our approach follows Refs.~\onlinecite{Sipe1993,Aversa1994,Aversa1995} in using the length gauge and separating the intraband and interband motion from the start. 
This avoids unphysical divergencies that plagued some earlier calculations.\cite{Sheik-Bahae1990,Sheik-Bahae1991,Sheik-Bahae1994,Moss1990}
The perturbative expansion, where intraband and interband motion are treated equivalently, yields macroscopic susceptibilities and ac conductivities, from which in turn one obtains nonlinear absorption coefficients. 
However, a main aspect of this work is the non-perturbative evaluation of the total absorption of (one or several) strong pulses or the absorption of a weak probe pulse in the presence of a strong pump pulse. 
Our method can be applied to arbitrary band structures. 
To demonstrate the basic principles we focus on a two-band model with $k$-independent dipole matrix element.
Due to their self-consistent calculation of the interband dipole matrix element, Aversa \textit{et al.}\cite{Aversa1994} obtained a different frequency dependence of the 2PA coefficient in an analogous model, but the general tendency of a steep increase towards strongly non-degenerate 2PA\cite{Cirloganu2011} is fully reflected by our work.
In other aspects we find good agreement with the literature on 2PA. 
More unexpected results are found for higher order MPA, which we evaluate analogously. In particular, the 3PA coefficient is found to vanish in the one-dimensional model if the frequency sum assumes a certain value not far above the band gap. The transients can be generally explained by pulse area overlaps as long as no saturation effects occur. Dynamic effects are discussed in the regime of high pump intensities.

Our paper is organized as follows.
The general formalism for obtaining nonlinear absorption rates from the SBE is outlined in Sec.~\ref{sec:formalism}. We are interested not only in absorption coefficients describing a perturbative steady state response, but also in the non-perturbative absorption of short and possibly strong pulses.
While the absorption of a single pulse is anticipated in Sec.~\ref{sec:absorptionrate}, the results for the pump-probe scheme are discussed in Sec.~\ref{sec:evaluation}.
Analytical 2PA coefficients are presented in Sec.~\ref{sec:2PA} explicitly for the two-band model and compared with previous works. 
In Sec.~\ref{sec:MPA} the SBE are evaluated to higher orders in the intensity of the pump pulse. 
We discuss both the change of third order effects (2PA) as well as the appearance of higher order effects (MPA) for more intense pump pulses. 
In principle this expansion can be taken to arbitrary order. In Sec.~\ref{sec:intensepump} we evaluate the approach including the pump pulse to infinite order.
In Sec.~\ref{sec:summary} we summarize our main results and discuss further directions.

\section{Theoretical formalism}
\label{sec:formalism}

In the following, we present the approach which we use to analyze the nonlinear optical response of bulk semiconductor and extended semiconductor nanostructures
(of dimension $d$) within a two-band model.
Following previous works we use the length gauge to describe the light-matter interaction including inter- and intraband excitations.
The resulting SBE have been employed previously to study, e.g.,
coherent effects induced by static and time-dependent electric fields in semiconductor superlattices \cite{Meier1994,Meier1995},
the generation of photocurrents by two-color fields \cite{Duc2005,Duc2006},
high-harmonic generation resulting from the excitation with intense optical \cite{Golde2008} and THz fields \cite{Golde2011},
as well as transient Wannier-Stark localization \cite{Schmidt2018}.

Here, we consider the case of a homogeneous excitation with several radiation fields of arbitrary frequencies.
Propagation effects as well as the many-body Coulomb interaction are neglected for simplicity
since we focus on the interplay between inter- and intraband excitations in particular for strongly non-resonant excitation conditions. 
Dephasing and relaxation processes are treated phenomenologically and sometimes disregarded in order to obtain analytical expressions.

We start by presenting the SBE and then describe the general perturbative expansion in powers of the optical field.
Next we consider a pump-probe set-up and include the propagation directions of the optical pulses into the analysis.
Furthermore, we present equations which allow to include the pump field non-perturbatively.

\subsection{Non-perturbative extended Bloch equations}

The system is described microscopically by a two-band model Hamiltonian
$\hat{H}=\hat{H}_0+\hat{H}_\mathrm{LM}$,
where $\hat{H}_0$ is the single-particle Hamiltonian of the valence and conduction band electrons,
$\hat{H}_0 = \hbar\sum_\mathbf{k} ( \omega_{v\mathbf{k}}\hat{a}_{v\mathbf{k}}^\dagger \hat{a}_{v\mathbf{k}}+\omega_{c\mathbf{k}} \hat{a}_{c\mathbf{k}}^\dagger \hat{a}_{c\mathbf{k}} )$.
The semiclassical light-matter interaction is given by $\hat{H}_\mathrm{LM}=\hat{H}_e+\hat{H}_i$, 
where in dipole approximation the interband part $\hat{H}_e$ and the intraband part $\hat{H}_i$ are given by 
\begin{align}
\hat{H}_e &= - \sum_\mathbf{k} \mathbf{E} \cdot (\mathbf{d}^* \hat{a}_{v\mathbf{k}}^\dagger \hat{a}_{c\mathbf{k}}+ \mathrm{h.c.}) ,\\
\hat{H}_i &= - \sum_{\mathbf{k},\mathbf{k}',\lambda=c,v} \langle \lambda,\mathbf{k}' | ie\mathbf{E}\cdot\nabla_\mathbf{k}|\lambda,\mathbf{k} \rangle \hat{a}_{\lambda\mathbf{k}'}^\dagger\hat{a}_{\lambda\mathbf{k}},
\end{align}
where $\mathbf{d}$ is the interband transition dipole matrix element which is taken as $k$-independent.

The Bloch equations are obtained from the Heisenberg equations of motion for the hole occupation
$n_{h\mathbf{k}}=1-\langle\hat{a}_{v\mathbf{k}}^\dagger \hat{a}_{v\mathbf{k}}\rangle$,
the electron occupation
$n_{e\mathbf{k}}=\langle\hat{a}_{c\mathbf{k}}^\dagger \hat{a}_{c\mathbf{k}}\rangle$,
and the interband polarization
$p_{\mathbf{k}}=\langle\hat{a}_{v\mathbf{k}}^\dagger \hat{a}_{c\mathbf{k}}\rangle$
and read \cite{Meier1994,Meier1995a,Duc2005} 
\begin{align}
\label{eq:p-eq}
\frac{\partial}{\partial t} {p}_\mathbf{k} &= -i\, ({\omega}_\mathbf{k}-i\gamma_p) p_\mathbf{k} + \frac{i}{\hbar}\, \mathbf{d}\cdot \mathbf{E}\, (1-n_{\mathbf{k}}) + \frac{e}{\hbar}\mathbf{E}\cdot \nabla_\mathbf{k}\, p_\mathbf{k}
 ,\\
\label{eq:ne-eq}
\frac{\partial}{\partial t} {n}_{q\mathbf{k}} &= \frac{i}{\hbar}\, \mathbf{d}\cdot \mathbf{E}\, (p_\mathbf{k}^*-p_\mathbf{k}) + \frac{e}{\hbar}\mathbf{E}\cdot \nabla_\mathbf{k}\, n_{q\mathbf{k}} - \gamma_n (n_{q\mathbf{k}}- n_{q\mathbf{k}}^{(0)}),
\end{align}
where the second equation holds for $q=e,h$ and we defined the transition frequency
${\omega}_\mathbf{k}\equiv\omega_{c\mathbf{k}}-\omega_{v\mathbf{k}}$ and introduced $n_{\mathbf{k}}\equiv n_{e\mathbf{k}}+n_{h\mathbf{k}}$, i.e.,
the sum of the electron and hole carrier occupations.
The relaxation and dephasing constants are denoted by $\gamma_n=1/T_1$ and $\gamma_p=1/T_2$, respectively.
Both evolution equations contain two types of source terms, an interband term (labeled $e$) and an intraband term (labeled $i$) which are treated equivalently in the following analysis.

\subsection{Perturbative expansion}

Macroscopic response functions (susceptibilities and conductivities) are obtained by expanding the quantities  $X_\mathbf{k}=\{p_\mathbf{k},n_{e\mathbf{k}},n_{h\mathbf{k}}\}$ in powers of the 
exciting field $\mathbf{E}(t)$ as $X_\mathbf{k}(t)=\sum_{m=0}^\infty X_\mathbf{k}^{(m)}(t)$ with
$X_\mathbf{k}^{(m)} \propto \mathbf{E}^m$. 
In our approach $n_{q\mathbf{k}}^{(0)}=n_{q\mathbf{k}}(t\to\pm\infty)$ acts at the same time as the initial density and as the equilibrium distribution, cf.~Eq.~(\ref{eq:ne-eq}).
Due to the intraband excitation, a finite initial density $n_{q\mathbf{k}}^{(0)}$ serves as a source for a linear electronic response
which contributes to the optical response in third order.  

\begin{figure}
	\includegraphics[width=.5\textwidth]{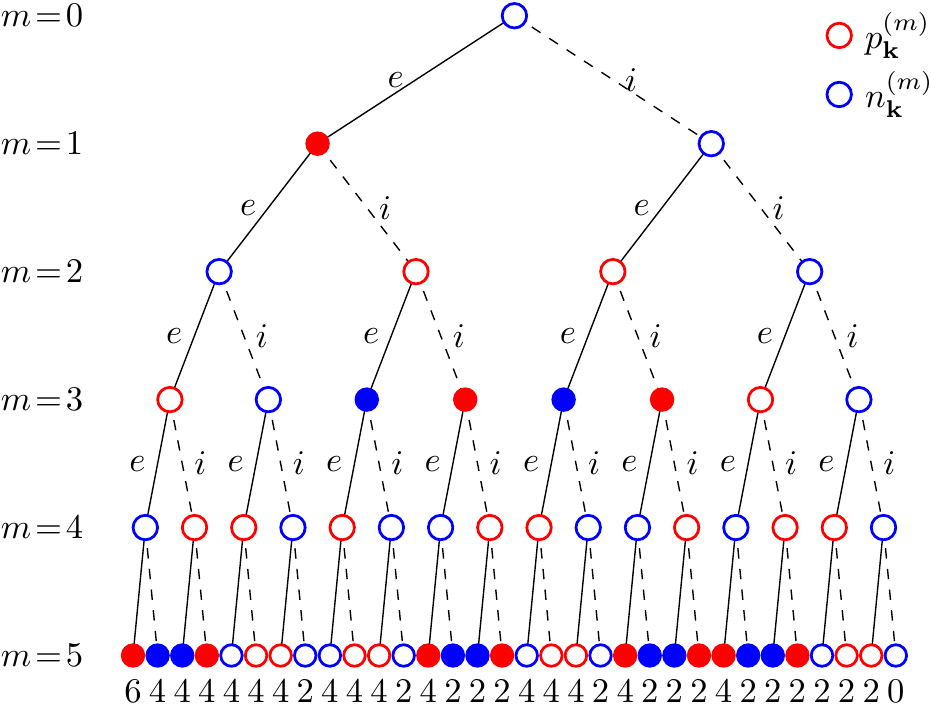}
	\caption{(color online) Schematics of the perturbative pathways up to fifth order (after Ref.\onlinecite{Aversa1995}). The full circles indicate paths contributing to the optical absorption of $(m+1)/2$ photons. The right half of the tree (intraband term for $m=1$) vanishes for an initially unexcited system.
The transition dipole matrix element $\mathbf{d}$ enters only through its projection onto the polarization directions of the incident fields.
We consider linearly parallel polarized pulses, i.e., ${\mathbf{e}}_1={\mathbf{e}}_2$ and thus introduce $\mu=\mathbf{d} \cdot \hat{\mathbf{e}}_1$.
The power of $\mu$ of each absorption contribution is shown below $m=5$-circles.
	}
	\label{fig:path-tree}
\end{figure}

In first-order in the optical field ($m=1$) the evolution equations are given by
\begin{align}
\label{eq:SBEexp1st}
\frac{\partial}{\partial t} {p}_\mathbf{k}^{(1)} &= \frac{i}{\hbar}\mathbf{d}\cdot \mathbf{E}\, \big(1-n_{\mathbf{k}}^\mathrm{(0)} \big) - i ({\omega}_\mathbf{k}-i\gamma_p)\, p_\mathbf{k}^{(1)}, \\
\frac{\partial}{\partial t} {n}_{q\mathbf{k}}^{(1)} &= \frac{e}{\hbar}\mathbf{E}\cdot \nabla_\mathbf{k}\, n_{q\mathbf{k}}^\mathrm{(0)} - \gamma_n n_{q\mathbf{k}}^{(1)}.
\end{align}
The $e$ and $i$ source terms lead to ${p}_\mathbf{k}^{(1)}$ and $n_{q\mathbf{k}}^{(1)}$, respectively.
Also for higher orders ($m>1$) both types of source terms appear in each equation
\begin{align}
\label{eq:SBEexpansion}
\frac{\partial}{\partial t} {p}_\mathbf{k}^{(m)} &= - \frac{i}{\hbar}\mathbf{d}\cdot \mathbf{E}\, n_{\mathbf{k}}^\mathrm{(m-1)} + \frac{e}{\hbar}\mathbf{E}\cdot \nabla_\mathbf{k}\, p_\mathbf{k}^{(m-1)} \nonumber\\
&\quad - i ({\omega}_\mathbf{k}-i\gamma_p) p_\mathbf{k}^{(m)}, \\
\frac{\partial}{\partial t} {n}_{q\mathbf{k}}^{(m)} &= \frac{i}{\hbar}\mathbf{d}\cdot \mathbf{E}\big(p_\mathbf{k}^{*(m-1)}\!-p_\mathbf{k}^{(m-1)}\big) + \frac{e}{\hbar}\mathbf{E}\cdot \nabla_\mathbf{k} n_{q\mathbf{k}}^{(m-1)} \nonumber\\
&\quad - \gamma_n n_{q\mathbf{k}}^{(m)}.
\label{eq:SBEexpansion_n}
\end{align}
Due to the two source terms in each equation, the total result in $m$th order is given by a sum of 2$^m$ terms involving all combinations of the different source terms\cite{Aversa1995}, cf.~Figure~\ref{fig:path-tree}.
The evolution equations for the individual pathways are given exemplary for up to $m=2$ in the Appendix~\ref{sec:paths}.
Since only some of these pathways are relevant for a particular nonlinear optical process, it is instructive to study each contribution separately by zeroing all others.
Below we label the pathways by a series of $e$'s and $i$'s, where the lowest order is given on the right.
For example, in second order there are two paths $ie$ and $ei$ leading to $p_\mathbf{k}^{(2)}$, as well as two paths $ee$ and $ii$ leading to $n_\mathbf{k}^{(2)}$. 
A path with an odd (even) number of $e$ terms leads to a polarization (density).
All paths starting from $i$ in first order vanish for a clean, cold semiconductor, i.e., without any electron and hole occupations  present prior to the optical excitation.\cite{Aversa1995} 

A truncation of the perturbative expansion of $X_\mathbf{k}(t)$, i.e., Eqs.~(\ref{eq:SBEexp1st})-(\ref{eq:SBEexpansion_n}),
requires that the incident electric fields are sufficiently weak such that $|X_\mathbf{k}(t)^{(m)}|\ll 1$ for all $m>0$.

\subsection{Absorption rate - general formalism}
\label{sec:absorptionrate}

The rate at which the energy of a light beam is absorbed per unit volume is generally given by a macroscopic spatial average of the product of the current density $\mathbf{j}(\mathbf{r},t)$ and the field amplitude $\mathbf{E}(\mathbf{r},t)$, i.e.,
$\langle \mathbf{j} \cdot \mathbf{E} \rangle$.\cite{Mahr1975}
This definition still holds in a non-perturbative regime, where the incident and transmitted pulse shapes generally differ.
Since propagation effects are disregarded here, i.e., we assume that variations of the light intensity within the sample are small, this averaging does not need to be carried out explicitly and we can set
$\mathbf{r}=0$.
For the case of two or more non-copropagating incident light beams, the spatial averaging is carried out as a directional expansion as described in Sec.~\ref{sec:pu-pr}.

The electric current density $\mathbf{j}$ is induced in the medium by $\mathbf{E}$ either via a change of the interband polarization or via carrier occupations, $\mathbf{j}=\mathbf{j}_p+\mathbf{j}_n$.\cite{Sipe1993}
For an isotropic band structure the part $\mathbf{j}_n$ vanishes if the densities are symmetric distributions in $k$-space. A contribution $\langle \mathbf{j}_n \cdot \mathbf{E} \rangle$ arises from density distributions which are asymmetric along the $k$-direction parallel to $\mathbf{E}$. For symmetric $n_{q\mathbf{k}}^{(0)}$ these contributions correspond to conductivity paths with an odd number of intraband terms $i$.
Transforming the $k$-sums into integrals, $\sum_\mathbf{k}\to (L/2\pi)^d\int_{\,\mathrm{BZ}}\!\!\mathrm{d}\mathbf{k}$, and omitting the prefactor, the two contributions to the current density are given by
\begin{align}
\mathbf{j}_p &=  \frac{\partial }{\partial t}   \mathbf{P}, \quad
 \mathbf{P}=  \mathbf{d}\int_{\mathrm{BZ}}\!\!\mathrm{d}\mathbf{k}\ (p_\mathbf{k}+p_\mathbf{k}^*) ,
\label{eq:j_p} \\
\mathbf{j}_n &= -e \int_{\mathrm{BZ}}\!\!\mathrm{d}\mathbf{k}\left( n_{e\mathbf{k}} \nabla_\mathbf{k} \omega_{c\mathbf{k}} - 
n_{h\mathbf{k}} \nabla_\mathbf{k} \omega_{v\mathbf{k}} \right) . \label{eq:j_n}
\end{align}
Substituting perturbative quantities $X_\mathbf{k}^{(m)}$ into these equations one analogously obtains perturbative results for the current density $\mathbf{j}^{(m)}$.

Inspecting the perturbative expansion of the Bloch equations we see that the excited densities are equal in both bands, $n_{e\mathbf{k}}-n_{e\mathbf{k}}^{(0)}=n_{h\mathbf{k}}-n_{h\mathbf{k}}^{(0)}$,
if the optically irrelevant contributions from paths without interband excitation are  neglected.
Thus, for our optical studies instead of Eq.~(\ref{eq:j_n}) we use
\begin{align}
\mathbf{j}_n &= - \frac{e}{2}\int_{\mathrm{BZ}}\!\!\mathrm{d}\mathbf{k}\ n_{\mathbf{k}} \nabla_\mathbf{k} \omega_{\mathbf{k}} ,
\label{eq:j_n_simpler}
\end{align}
and have to solve the $n$-equations not individually for each band but only as a sum. 

Here we describe the two-band models used in this work. 
As long as no initial densities $n_{q\mathbf{k}}^{(0)}$ are present,
only the valence to conduction interband transition frequency $\omega_{\mathbf{k}}$
enters the Bloch equations along with Eqs.~\eqref{eq:j_p},\eqref{eq:j_n_simpler}
and the individual band dispersions are not relevant.
We either use a one-dimensional tight-binding band structure with lattice constant $a$,
\begin{equation}\label{eq:TBmodel}
\omega_k = \omega_\mathrm{g} + \frac{\omega_\mathrm{b}}{2} [1-\cos(k a) ] ,
\end{equation}
where $\hbar \omega_\mathrm{g}$ is the band gap and $\hbar \omega_\mathrm{b}$ the joint band width.
This model is particularly suitable for very strong fields where the electronic excitations are driven through the entire Brillouin zone.
The other model used for analyzing the optical response near the band gap is a parabolic dispersion
\begin{equation}\label{eq:parabolic}
\omega_\mathbf{k}=\omega_\mathrm{g}+\hbar\mathbf{k}^2/2m^*,
\end{equation}
where $1/m^* = 1/m_e^* + 1/m_h^*$ is the inverse reduced effective mass. 
When considering initial quasiequilibrium carrier distributions $n_{q\mathbf{k}}^{(0)}$, the individual effective masses ($m_e^*,m_h^*$) need to be specified. Here we choose the values for the lowest conduction and the heavy hole bands of GaAs. 

\begin{figure}
	\includegraphics[width=.5\textwidth]{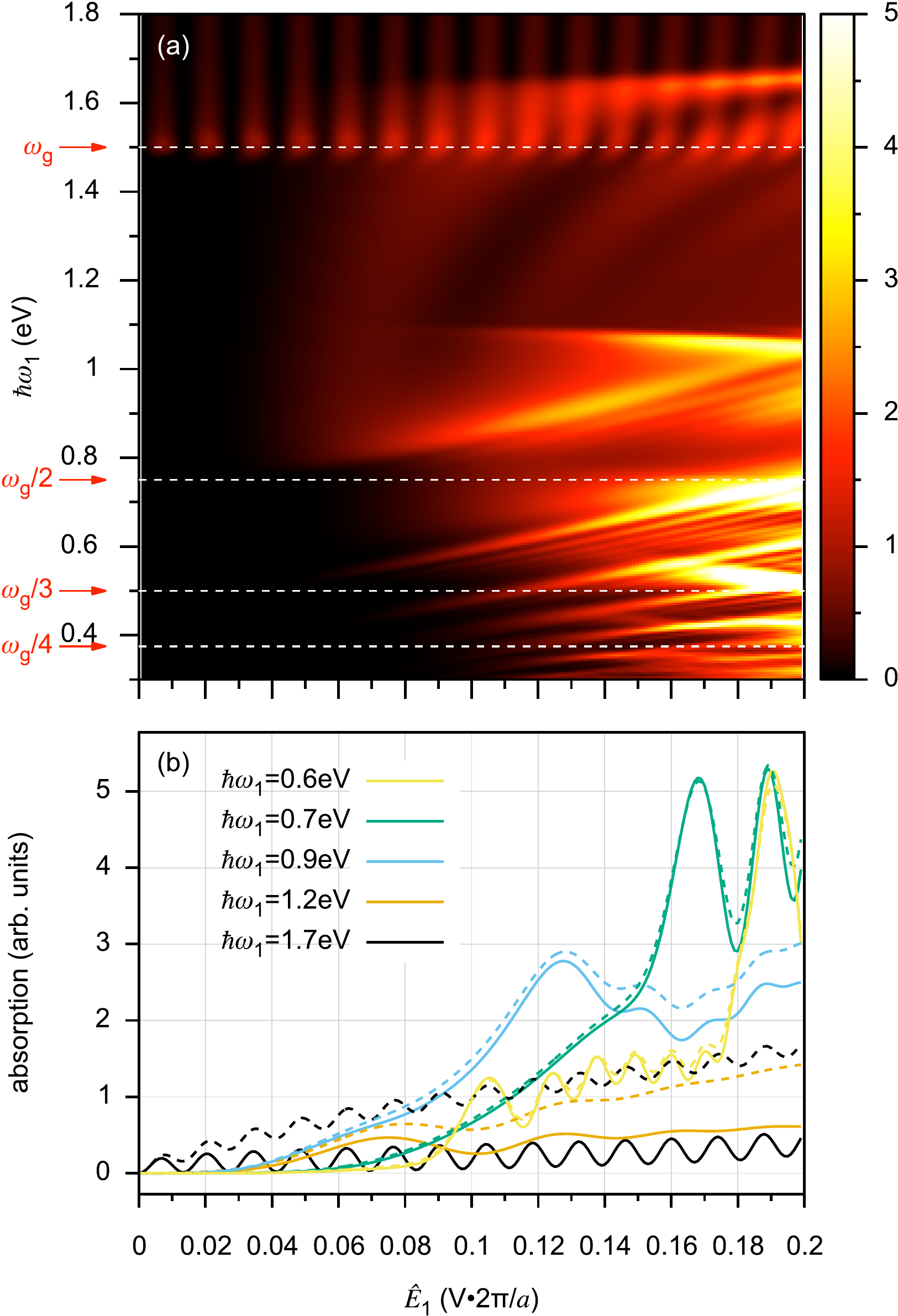}
	\caption{(color online) (a) Absorption (arbitrary units) of a single Gaussian pulse of frequency $\omega_1$, peak amplitude $\hat{E}_1$,
		and duration $\tau_1=50$~fs by a one-dimensional sample using the tight-binding model \eqref{eq:TBmodel} with $\hbar\omega_\mathrm{g}=$1.5~eV, $\hbar\omega_\mathrm{b}=$1.3~eV.
		The dephasing time $T_2$ is infinite. The colorscale does not include all data values, but the off-range regions ($>5$) are small. 
		(b) Solid lines are the same data as in (a), the additional dashed lines are for $T_2=$200~fs.
	}
	\label{fig:1col_abs}
\end{figure}

Further simplifications arise since we consider linearly and parallel polarized incident pulses here.
In this case the transition dipole $\mathbf{d}$ enters only through its projection $\mu$ onto the field direction. 
Here we use a constant, i.e., $k$-independent, interband dipole matrix element suitable for bulk GaAs of
$\mu=3e a/5.65$, that is $\mu=3e$\AA~ after substituting the lattice constant of GaAs, $a \approx $ 5.65\AA.
The lattice constant $a$ does not enter our calculations explicitly, but is transferred to other quantities such as the field strength in Figure~\ref{fig:1col_abs}.

Before we describe the generalization of the light absorption formalism to multiple propagation directions in the next Section, we evaluate here the approach for a single pulse.
Besides serving as a simple example, this is experimentally relevant since degenerate MPA can be measured this way.
We assume an initially unexcited, one-dimensional sample which is described by a two-band tight-binding band structure \eqref{eq:TBmodel}.
Figure~\ref{fig:1col_abs}(a) and (b) show the non-perturbative absorption, calculated as the time integral of $j E$, of a single pulse with a Gaussian envelope
$e^{-t^2/\tau_1^2}$ and different central frequencies.

The one-photon absorption (1PA) resonant region, $\omega_1>\omega_\mathrm{g}$, shows regular Rabi oscillations of period $\Delta\hat{E}_1=\hbar/ (\sqrt{\pi}\mu\tau_1) \approx 0.014$V$\cdot 2\pi/a$.
For weak fields, $\hat{E}_1\ll\Delta\hat{E}_1$, the 1PA increases quadratically with the field amplitude $\hat{E}_1$ as expected from perturbation theory.
Figure~\ref{fig:1col_abs}(b) shows explicitly for $\hbar\omega_1\approx 1.7$eV that the absorption increases roughly linearly with each Rabi cycle in the case $T_2\to\infty$. For finite $T_2$ the increase is stronger and follows a  $\hat{E}_1^{1/2}$-dependence when the excitation frequency is within the parabolic region of the band structure.
As expected, the effect of the intraband motion is negligible for 1PA case in the considered field amplitude range. 

In the frequency region $\omega_\mathrm{g}/2<\omega_1<(\omega_\mathrm{g}+\omega_\mathrm{b})/2$, which partly overlaps with other MPA regions, 2PA can be clearly recognized.
Since 2PA scales as $\propto\hat{E}_1^4$ at field amplitudes well below saturation effects, it is negligible compared to 1PA for weak fields. 
Saturation sets in at $\hat{E}_1 \approx 0.05$V$\cdot 2\pi/a$, corresponding to a peak intensity of 41 GW/cm$^2$ (using the lattice constant of GaAs).
This agrees in order of magnitude with the 2PA saturation observed in bulk\cite{Lami1996} and monolayer\cite{Dong2018} semiconductors. 
We note that Rabi-like oscillations caused by 2PA appear in two different ways. 
For higher frequencies, e.g.~$\hbar\omega_1\approx 1.2$eV, weak modulations of large period are visible. On the other hand, more regular Rabi-modulations with shorter period appear beyond a certain resonance signal which starts slightly above $\omega_\mathrm{g}/2$ (the 2PA resonance with the lower band edge) and is Stark shifted to higher frequencies. The shift is first quadratic (at about $\hat{E}_1 \approx 0.06$V$\cdot 2\pi/a$) and later linear in $\hat{E}_1$. 
Below of this resonance line further lines appear regularly and the entire absorption edge remains roughly at $\omega_\mathrm{g}/2$ (hard to distinguish due to the overlap of 2PA and 3PA). 
Generally, Rabi-like modulations for 2PA appear to be weaker than for 1PA (due to the effect of intraband motion). This may be the reason why two-photon Rabi-cycles have been observed
so far only in quasi zero-dimensional systems, such as atoms,\cite{Gentile1989} molecules,\cite{Linskens1997} and semiconductor quantum dots.\cite{Stufler2006}

For 3PA the behavior is somewhat analogous and even more easily distinguishable. The most dominant signal is the band gap resonance $\omega_\mathrm{g}/3$, which is again Stark shifted upwards. 
In the perturbative analysis in Sec.~\ref{sec:MPA} we will discuss in more detail that this dominant signal is separated from a broader 3PA range further above the band gap, which is also visible in Figure~\ref{fig:1col_abs}(a). 
Relatively pronounced Rabi oscillations appear beyond this upwards shifting band gap resonance. 
The field amplitude range where perturbative analysis is valid for 3PA is very small since the weak field regime for 3PA with a scaling $\propto\hat{E}_1^6$
has only a narrow validity range before also higher orders contribute.
Note also that all higher MPA sets in at roughly comparable field amplitudes.
Another strong 3PA signal is the resonance with the upper band edge $\omega_\mathrm{g}+\omega_\mathrm{b}$.

For higher MPA basically only narrow resonance lines appear, which are resonant with frequencies slightly above the band gap.
Here we mainly see the linear frequency shift of these resonances. We confirm   in the perturbative analysis in Sec.~\ref{sec:MPA}
that 4PA is indeed restricted to a rather narrow range above the band gap.
In our model, for certain MPA, e.g.~6PA, again another signal appears corresponding to a resonance with the upper band edge and Stark-shifting to lower frequencies
for strong fields.

\subsection{Perturbative treatment of pump-probe experiments}
\label{sec:pu-pr}

We consider MPA as measured via the differential absorption of a probe pulse ('1') in the presence of a non-copropagating pump pulse ('2').
The probe pulse is assumed to be sufficiently weak so that material excitation of higher than linear order in its field amplitude can be neglected.
Depending on the pump intensity $I_2$, nonlinear-optical processes consisting of the simultaneous absorption of one probe photon and $(l-1)$ pump photons are relevant, which are described by a $l$PA-coefficient $\alpha^{(l)}$. 
The total absorption coefficient is thus expanded in the pump intensity $I_2$ as
$\alpha = I_1 \sum_{l=1}^\infty \alpha^{(l)} I_2^{l-1}$.
The formalism presented here is also applicable without the restriction to weak probe pulses or to the number of pulses, but for such cases the number of required equations in the directional expansion, see below, grows strongly.

The different propagation directions of pump and probe are taken into account by including the dependence of the field on the macroscopic position vector $\mathbf{r}$.\cite{Binder1991}
The individual fields are plane waves with a temporal envelope and a central frequency
$\mathbf{E}_\nu=\hat{\mathbf{e}}_\nu\hat{E}_\nu(t) \cos(\mathbf{k}_\nu \cdot \mathbf{r}-\omega_\nu t)$, for $\nu=1,2$.
We assume the two fields to have the same linear polarization, i.e., $\hat{\mathbf{e}}_1=\hat{\mathbf{e}}_2$.
As mentioned above, the transition dipole matrix element $\mathbf{d}$ enters only through its projection onto the field direction, $\mu=\mathbf{d} \cdot \hat{\mathbf{e}}_1$.
The intensities are obtained from the field amplitudes as
$I_\nu(t)=(\epsilon_0 c/2) |\hat{E}_\nu(t)|^2$.
In this work we consider either $\hat{E}_\nu(t) = \hat{E}_\nu =\ $const. (continuous wave limit) or finite pulses with a Gaussian envelope function $\hat{E}_\nu(t)$.
Using a standard perturbative expansion in the two field amplitudes $\mathbf{E}_\nu$ (and abbreviating $\mathbf{j}^{(m)}\equiv \mathbf{j}^{(m)(1|m-1)}$
where in the latter expression the superscripts $1$ and $m-1$ denote the orders in the probe and pump field amplitude, respectively), 
we find the MPA coefficient\cite{Mahr1975} in the steady-state response as 
\begin{align}
\label{eq:PrAbs}
\alpha^{(l)}(\omega_1,\omega_2) &=\frac{\langle \mathbf{j}^{(2l-1)} \cdot \mathbf{E}_1 \rangle}{2 l I_1 I_2^{l-1}}.
\end{align} 
The factor $l$ in the denominator ensures that for $\omega_1=\omega_2$ the obtained absorption coefficient agrees with the single beam setup, i.e., $\alpha^{(l)}(\omega_1,\omega_1)=\alpha^{(l)}(\omega_1)$.
For pulses of finite duration we have accordingly
\begin{align}
\label{eq:PrAbs_pulse}
\alpha^{(l)}(\omega_1,\omega_2) &=\left(\frac{2}{\epsilon_0 c}\right)^l\frac{\int \mathrm{d}t\,
	 \langle\mathbf{j}^{(2l-1)}(t) \cdot \mathbf{E}_1(t)\rangle}{2l\int \mathrm{d}t \hat{E}_1^2(t) \hat{E}_2^{2l-2}(t)}.
\end{align}
Due to the spectral width of the pulses, this expression yields approximately the frequency dependence of the absorption coefficient. 

Here we replace the spatial averaging appearing in Eqs.~\eqref{eq:PrAbs} and \eqref{eq:PrAbs_pulse} by a directional expansion. 
The directional expansion differs from the power expansion only in the splitting of each field into its two Fourier components,
$\cos(\mathbf{k}_\nu \cdot \mathbf{r}-\omega_\nu t) \to \frac{1}{2} ( \exp[i(\mathbf{k}_\nu \cdot \mathbf{r}-\omega_\nu t)] + \exp[- i(\mathbf{k}_\nu \cdot \mathbf{r}-\omega_\nu t)]) $,
corresponding to forward/backward propagating complex fields. 
The current density $\mathbf{j}^{(m)(a|b)}$ is obtained from the complex quantities $p_{\mathbf{k}}^{(m)(a|b)}$ and $n_{q\mathbf{k}}^{(m)(a|b)}$,
where as before the superscript $(m)$ refers to the total order and the additional superscript $(a|b)$ denotes that the term is proportional to $\exp[i( a \mathbf{k}_1 + b \mathbf{k}_2) \cdot \mathbf{r}]$, i.e., $(1|0)$ corresponds to the probe direction.\cite{Lindberg1992,Sieh1999,Weiser2000,Binder1991}
In the steady-state response the probe-directed signals are related to the field amplitudes, e.g., in third order, by
\begin{align}
\label{eq:chi}
\hat{\mathbf{e}}_1 \cdot \mathbf{P}^{(3)(1|0)} &= \chi^{(3)}(\omega_1;\omega_1,\omega_2,-\omega_2) \hat{E}_1 \hat{E}_2 \hat{E}_2^* e^{i(\mathbf{k}_1  \cdot \mathbf{r}-\omega_1 t)},\\
\label{eq:sigma}
\hat{\mathbf{e}}_1 \cdot \mathbf{j}_n^{(3)(1|0)} &= \sigma^{(3)}(\omega_1;\omega_1,\omega_2,-\omega_2) \hat{E}_1 \hat{E}_2 \hat{E}_2^* e^{i(\mathbf{k}_1 \cdot \mathbf{r}-\omega_1 t)},
\end{align}
where $\chi$ is the susceptibility and  $\sigma$ the ac-conductivity.
Here, we do not study the tensor character of the response functions since we consider linearly and parallel polarized fields and an isotropic medium, thus only the diagonal elements are relevant.
Using relations of the type \eqref{eq:chi}-\eqref{eq:sigma} in Eq.~(\ref{eq:PrAbs}) we obtain
\begin{align}
\label{eq:2PA}
\alpha^{(l)}(\omega_1,\omega_2)=
-\frac{1}{2}\left(\frac{2}{\epsilon_0 c}\right)^l
&\big[\mu\omega_1 \chi''^{(2l-1)} 
+\sigma'^{(2l-1)}\big],
\end{align}
where single (double) prime stands for the real (imaginary) part, 
and the arguments of $\chi$ and $\sigma$ are
$(\omega_1;\omega_1,\omega_2,-\omega_2,\ldots)$ according to the order $l$.

Note that a directional expansion may not be necessary for certain choices of parameters. For instance, in the case of non-degenerate 2PA the time integration in Eq.~\eqref{eq:PrAbs_pulse} automatically singles out the probe-directed contributions for sufficiently long pulses. 
The absorption coefficient for 2PA (\ref{eq:2PA}) was given by Bredikhin and Genkin,\cite{Bredikhin1970} 
but without the conductivity term, which plays an important role for $l>1$ even for the case of filled bands.
Third-order expressions for the 2PA coefficient for varying excitation frequencies were obtained using a transition rate approach\cite{Sheik-Bahae1994} and
by a susceptibility approach \cite{Aversa1994} and we compare to these results in Sec.~\ref{sec:2PA}.
Furthermore, the optical injection of photocurrents by two light beams having frequencies $\omega$ and $2\omega$ with $2 \hbar \omega$ larger than the gap energy
was studied by Atanasov \textit{et al}.\cite{Atanasov1996}.

\subsection{Bloch equations non-perturbative in the pump field}

So far we have treated both probe and pump field perturbatively, yielding absorption coefficients for each order of the pump intensity. 
When a distinction between these orders is not required or when the pump field is so intense that an order expansion cannot be applied,
it is possible to include the pump field non-perturbatively. 
We still consider a weak probe pulse so that its absorption remains linear in the probe intensity.
If both pulses are treated non-perturbatively, the directional expansion is inevitable and immensely involved.
As discussed below and in Sec.~\ref{sec:intensepump}, a directional expansion may be omitted for the weak probe scheme as long as the pump pulse intensity does not reach very high values.

The differential probe absorption in the presence of an arbitrarily strong pump field is extracted from the mixed quantities $\bar{p}_\mathbf{k},\bar{n}_{q\mathbf{k}}$ in the following set of SBE:\cite{Schmidt2018}
\begin{alignat}{2}
\label{eq:mixSBE_first}
\frac{\partial  }{\partial t} p_\mathbf{k} &= -i\, \bar{\omega}_\mathbf{k}\, p_\mathbf{k} + \frac{i}{\hbar}\,  \mathbf{d}\cdot \mathbf{E}_2\, (1-n_{\mathbf{k}}) + \frac{e}{\hbar} \mathbf{E}_2\cdot \nabla_\mathbf{k}\, p_\mathbf{k} , \\
\frac{\partial  }{\partial t} n_{q\mathbf{k}} &= - \gamma_n n_{q\mathbf{k}} + \frac{i}{\hbar} \mathbf{d}\cdot \mathbf{E}_2\, (p_\mathbf{k}^*-p_\mathbf{k}) + \frac{e}{\hbar} \mathbf{E}_2\cdot \nabla_\mathbf{k}\, n_{q\mathbf{k}}  , \\
\frac{\partial  }{\partial t}{p}_\mathbf{k}^{(1)} &= - i \bar{\omega}_\mathbf{k}\, p_\mathbf{k}^{(1)} + \frac{i}{\hbar} \mathbf{d}\cdot \mathbf{E}_1\, \big(1-n_{\mathbf{k}}^\mathrm{(0)} \big) , \\
\frac{\partial  }{\partial t}{\bar{p}}_\mathbf{k} &= -i\, \bar{\omega}_\mathbf{k}\, \bar{p}_\mathbf{k} - \frac{i}{\hbar} \mathbf{d}\cdot \mathbf{E}_1\, n_\mathbf{k} - \frac{i}{\hbar} \mathbf{d}\cdot \mathbf{E}_2\, \bar{n}_\mathbf{k} \\
&\quad + \frac{e}{\hbar} \mathbf{E}_1\cdot \nabla_\mathbf{k}\, p_\mathbf{k} + \frac{e}{\hbar} \mathbf{E}_2\cdot \nabla_\mathbf{k}\, ({p}_\mathbf{k}^{(1)}+ \bar{p}_\mathbf{k} ) , \nonumber \\	
\frac{\partial  }{\partial t}{\bar{n}}_{q\mathbf{k}} &= -\gamma_n \,\bar{n}_{q\mathbf{k}} + \frac{i}{\hbar} \mathbf{d}\cdot \mathbf{E}_1\,(p_\mathbf{k}^*-p_\mathbf{k}) 
+ \frac{e}{\hbar} \mathbf{E}_1\cdot \nabla_\mathbf{k}\, n_{q\mathbf{k}} \label{eq:mixSBE_last}\\
&\quad + \frac{e}{\hbar} \mathbf{E}_2\cdot \nabla_\mathbf{k}\, \bar{n}_{q\mathbf{k}} + \frac{i}{\hbar} \mathbf{d}\cdot \mathbf{E}_2\,(\bar{p}_\mathbf{k}^*-\bar{p}_\mathbf{k}+{p}_\mathbf{k}^{*(1)}-{p}_\mathbf{k}^{(1)} ) , \nonumber
\end{alignat}	
with $\bar{\omega}_\mathbf{k} \equiv \omega_{\mathbf{k}} - i \gamma_p$.
The quantities $p_\mathbf{k},n_{q\mathbf{k}}$ are purely due to the pump, while $p_\mathbf{k}^{(1)}$ is the linear probe polarization.
A possibly present initial density $n_{q\mathbf{k}}^\mathrm{(0)}$ needs also to be included by the initial condition $n_{q\mathbf{k}}(t \rightarrow -\infty)=n_{q\mathbf{k}}^\mathrm{(0)}$
whereas the mixed densities that are induced by both pulses start with $\bar{n}_{q\mathbf{k}}(t \rightarrow -\infty)=0$.
For sufficiently strong pump pulses, frequency components will appear near the probe frequency and other directions than the probe may contribute to the absorption obtained analogously to Eq.~(\ref{eq:PrAbs_pulse}). For weaker or longer pulses, such that only a maximum of $m$ orders is relevant, the response described by 
Eqs.~(\ref{eq:mixSBE_first})-(\ref{eq:mixSBE_last}) agrees with the perturbative expansion.

\section{Results for pump-probe scheme}
\label{sec:evaluation}

In the following we present and discuss our analytical and numerical results.
Various regimes of pump intensities are discussed successively, starting in Sec.~\ref{sec:2PA} from sufficiently weak pump intensities and an off-resonant probe such that a third-order analysis is valid.
Higher-order effects are evaluated for $m=5$ and $m=7$ in Sec.~\ref{sec:MPA}.
We analyze both the corresponding MPA coefficients ($\alpha^{(3)}$ and $\alpha^{(4)}$) and the corrections to the lower orders, e.g.~$\chi^{(5)}$-corrections for a resonant sum frequency, i.e., $\omega_1+\omega_2$ larger than the gap frequency.
Finally, the regime of very high pump intensities leading to dynamic saturation effects is investigated in Sec.~\ref{sec:intensepump}.

\subsection{Two-photon absorption in third order}
\label{sec:2PA}

Here we start by deriving analytical expressions for the 2PA coefficients in third order
in the continuous wave limit and compare them with numerical results obtained for Gaussian pulses. 
The susceptibility $\chi^{(3)}$ and the conductivity $\sigma^{(3)}$ both have the following form
\begin{align}
\label{chi3_pp}
\chi^{(3)}(\omega_1,\omega_2,-\omega_2) &= \int_{\mathrm{BZ}}\!\!\!\!\mathrm{d}\mathbf{k} \sum_{\mathrm{paths}} \sum_{\ell=1}^3 \sum_{\eta=\pm} \phi_\mathbf{k}^{(\mathrm{path};\ell)}(\omega_1,\eta\omega_2) ,
\end{align}
where the paths leading to $\chi^{(3)}$ are $\{iie$, $eee$, $iei$, $eii\}$, 
while those for $\sigma^{(3)}$ are $\{eie$, $iee$, $eei$, $iii\}$.
The two sums over subpaths ($\ell$ and $\eta$) are due to the order in which the fields $E_1,\ E_2,\ E_2^*$ appear in the perturbation expansion.
$\ell$ labels the order in which field $E_1$ appears, while $\eta$ labels whether $E_2$ or $E_2^*$ appears first in the series. 

\begin{figure}
	\includegraphics[width=.48\textwidth]{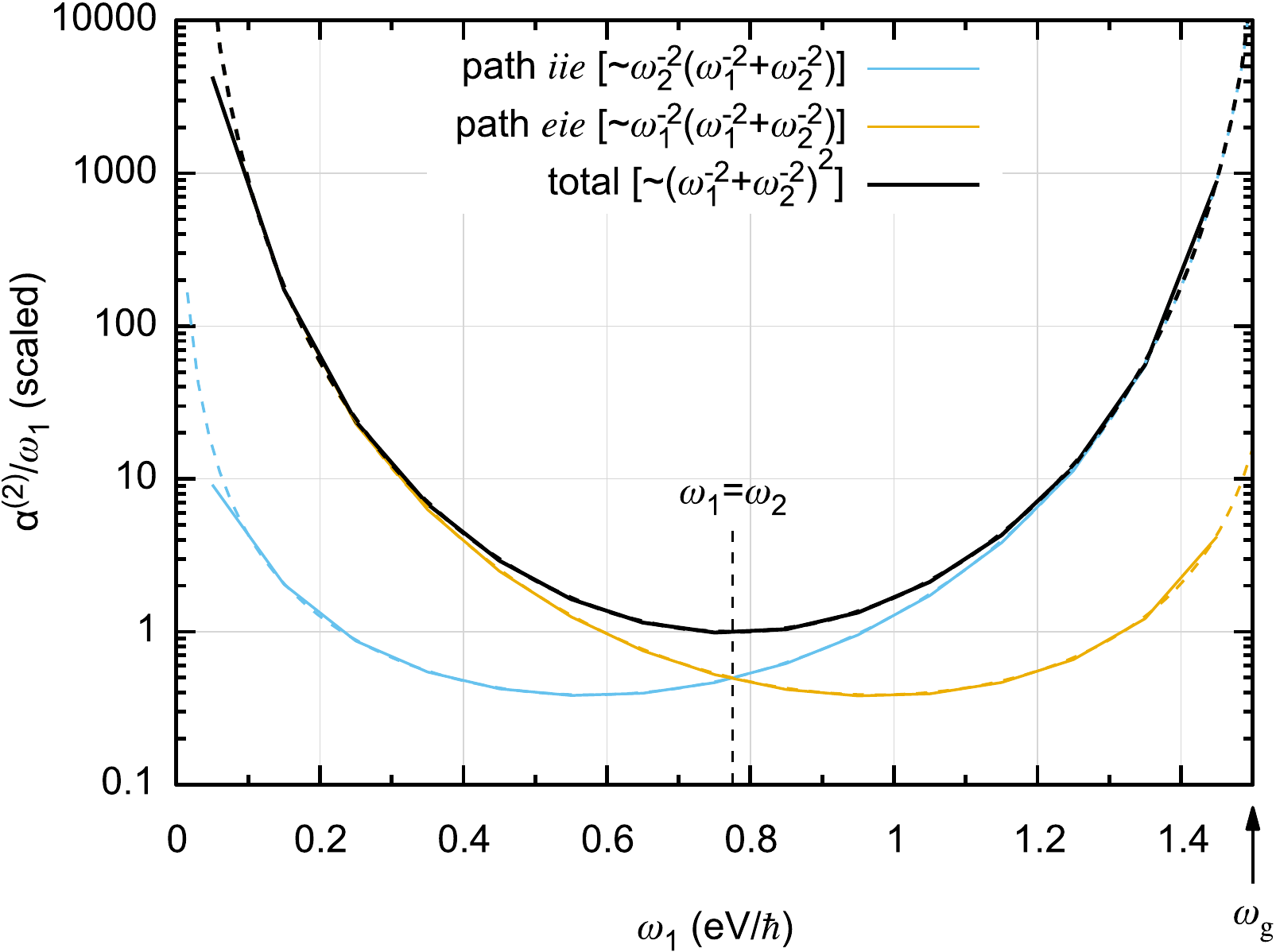}
	\caption{(color online) 2PA coefficient for different frequency ratios $\omega_1/\omega_2$ (three-dimensional model). The frequency sum $\omega_1+\omega_2=1.55$~eV$=\omega_\mathrm{g}+50$~meV is kept constant and $\omega_1$ is varied.
	Numerical results are shown by solid lines for $\tau_1=\tau_2=50$~fs, $T_1\to\infty$, and $T_2=200$~fs. 
			The dashed lines display the scaled analytical expressions given in the brackets, cf.~Eq.~(\ref{eq:2PA_1D}).
	}
	\label{fig:2PA_freq}
\end{figure}

From the analytical expressions for  $\phi_\mathbf{k}^{(\mathrm{path};\ell)}(\omega_1,\omega_2)$, see Appendix~\ref{app:phi3},
all contributing third-order effects, i.e., 2PA, electronic Raman, and optical Stark effect, can be deduced.
Compact results are obtained in the limit of vanishing relaxation and dephasing constants, i.e., $\gamma_n\to 0$ and $\gamma_p\to 0$.
Strictly speaking, the results depend on the order of the limits and we take $\gamma_n\to 0$ first
since the relaxation time is typically longer than the dephasing time.  
 
We restrict the analytical evaluation to the continuous wave limit to 2PA, hence we consider $0<\omega_\nu<\omega_\mathbf{k}$ for $\nu=1,2$.
In third-order four out of eight paths (indicated by solid circles in Figure~\ref{fig:path-tree}) show a resonance factor $(\omega_1+\omega_2-\bar{\omega}_\mathbf{k})^{-1}$ and are thus relevant for 2PA.
The other four paths, for which the second-order quantity is a density $n_\mathbf{k}^{(2)}$, actually lead to divergent absorption contributions, arising from a forbidden truncation of the perturbation series for continuous-wave fields and $\gamma_n=0$ and $\gamma_p\to 0$. 
Such contributions are thus omitted in our evaluations. For sufficiently short pulses, such these $n_\mathbf{k}^{(2)}$ remain small, the truncation of the perturbation series is allowed and these paths do not significantly contribute as confirmed by our numerical solutions. 
The 2PA coefficient scales as $\mu^2$ if only ,,contributing'' paths are considered - this is in contrast to the 3PA coefficient, where different powers of $\mu$ appear as indicated in Figure~\ref{fig:2PA_freq}.
It is noteworthy that the path $iei$ ($eei$), which vanishes for initially filled bands, is completely canceled by other terms of the path $iie$ ($eie$).
As a consequence the derivatives of the initial density, e.g.,~$\nabla_\mathbf{k} n_{k}^\mathrm{(0)}$, do not enter the 2PA coefficient. 

\begin{figure}
	\includegraphics[width=.49\textwidth]{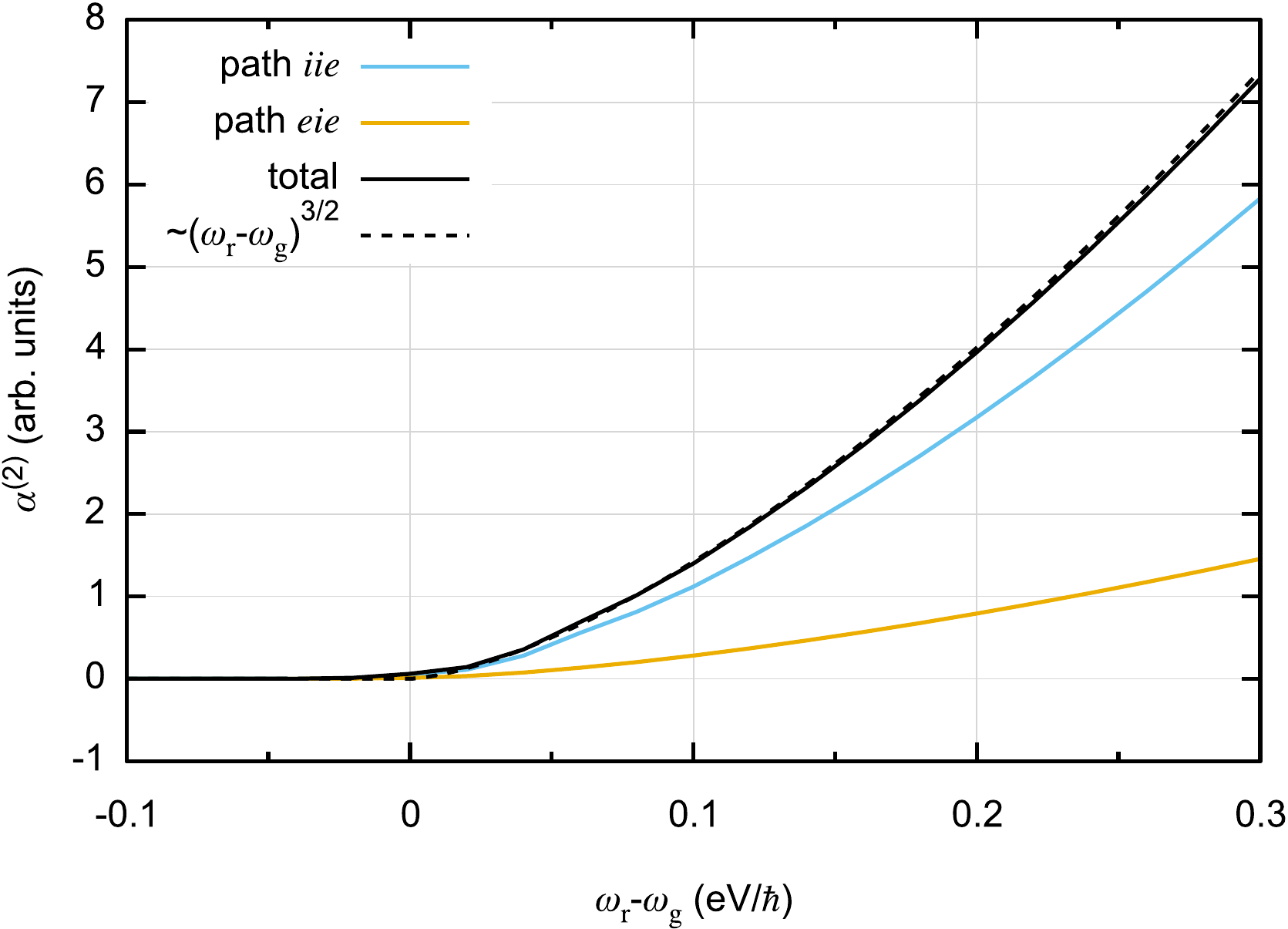}
	\caption{(color online) 2PA coefficient for varying band gap $\omega_\mathrm{g}$ in three dimensions. Numerical results (solid lines) are computed for two Gaussian pulses with
	$\tau_1=\tau_2=50$~fs; $\omega_1=2\omega_2=1$~eV/$\hbar$. 
		The dashed line is the fitted analytical expression from Eq.~\eqref{eq:bandgap-scaling}.
	}
	\label{fig:2PA_ogap}
\end{figure}

\begin{figure}[t]
	\includegraphics[width=.49\textwidth]{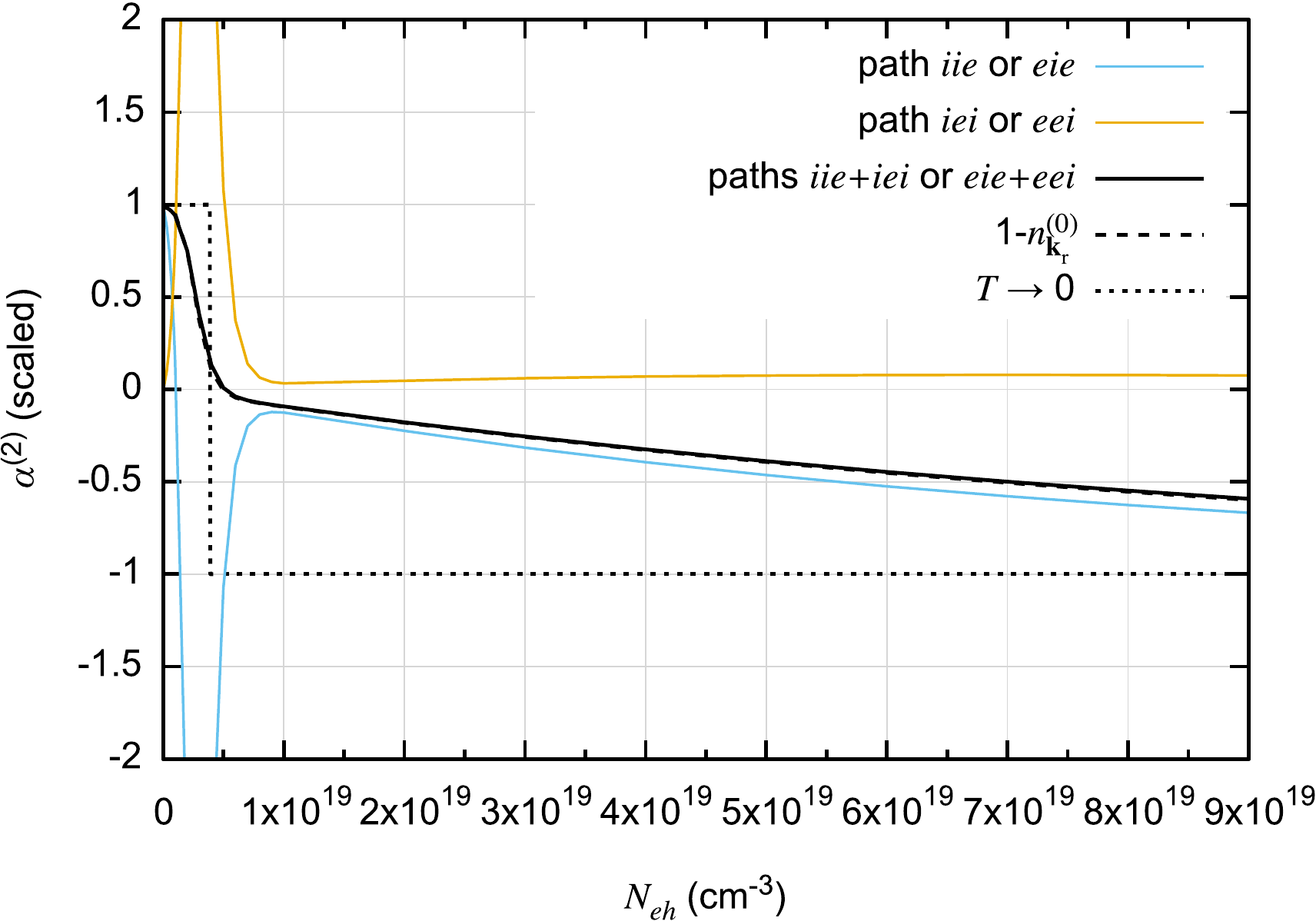}
	\caption{(color online) 2PA coefficient for two Gaussian pulses as function of an initial quasi-equilibrium carrier density in three dimensions.
	The considered parameters are $\omega_1+\omega_2=1.7$~eV, $\omega_1/\omega_2=2$, $\tau_1=\tau_2=50$~fs, $T_1\to\infty$, $T_2=200$~fs, and $T=300$~K.
The total of all paths (shown by the black curve) agrees well with the factor $1-n_e-n_h$ as expected from Eq.~\eqref{eq:2PA_1D}.
	For $T\to 0$ this factor approaches the step function shown by the dotted line.}
	\label{fig:N0--2PA}
\end{figure}

Using the response functions and the recipe outlined in Appendix~\ref{app:phi3}, the 2PA coefficient for a one-dimensional system is found from Eqs.~(\ref{eq:2PA}) and (\ref{chi3_pp}) in the limit $\gamma_n\to 0$ and $\gamma_p\to 0$ as 
\begin{align}
\label{eq:2PA_1D}
\alpha^{(2)}(\omega_1,\omega_2) &= 
\frac{2^3e^2\mu^2}{\hbar^3\epsilon_0^3c^2} \omega_1 (\omega_1^{-2}+\omega_2^{-2})^2
\big(1-n_{\mathbf{k}_\mathrm{r}}^\mathrm{(0)} \big)
 |v_{k_\mathrm{r}}| ,
\end{align}
where $k_\mathrm{r}$ is the positive resonant point in the first Brillouin zone, i.e.,
$\omega_{k_\mathrm{r}}=\omega_1+\omega_2$
 and $v_{k_\mathrm{r}}=\mathrm{d}\omega_k/\mathrm{d}k|_{k=k_\mathrm{r}}$ is the group velocity at this point,
where for simplicity we assume that the band structure is isotropic.
The dependencies of $\alpha^{(2)}$ on the frequency ratio $\omega_1/\omega_2$, the band structure, and initial band fillings are
independent of each other and will be discussed separately in the following.
Numerical results using Gaussian pulses of duration $\tau_\nu$ (solid lines) are compared with analytical continuous-wave expressions (dashed lines) in Figures~\ref{fig:2PA_freq}-\ref{fig:N0--2PA}.

Figure~\ref{fig:2PA_freq} shows the probe frequency dependence of the absorption for a fixed sum frequency $\omega_1+\omega_2$. Since the sample is initially unexcited,
only one susceptibility path ($iie$) and one conductivity path ($eie$) contribute (with 2 of 6 subpaths for each path).
All other paths are negligible, i.e., they vanish for the investigated analytical limit and are very small in the numerical evaluations with Gaussian pulses.
Nearly perfect agreement between numerical and analytical results is obtained as long as both dephasing and relaxation times are longer than the chosen pulse durations $\tau_\nu$.

We note that since each beam delivers one photon to the 2PA process, $\alpha^{(2)}(\omega_1,\omega_2)/\omega_1$ should be symmetric with respect to the two frequencies.
For weak relaxation this symmetry is fulfilled automatically. 
When we consider very short dephasing times in our numerical evaluations the absorption is reduced but still remains symmetric. 
On the other hand, since the relaxation time enters only in the conductivity path $eie$ the symmetry is lost for $T_1<\tau_\nu$.
The absorption coefficient diverges if one of the light frequencies vanishes and the other is resonant.
This drastic enhancement for non-degenerate excitation was also obtained previously.\cite{Sheik-Bahae1994,Aversa1994}

The general behavior of the 2PA coefficient for varying excitation frequencies is in good qualitative agreement with previous studies.
However, regarding the exact dependence differing results exist in the literature.
The frequency dependence obtained by the transition rate approach in Ref.~\onlinecite{Sheik-Bahae1994} is $\alpha^{(2)}(\omega_1,\omega_2)\propto(\omega_1+\omega_2)^2/\omega_1^3\omega_2^4$.
Aversa et al.\cite{Aversa1994} applied a susceptibility approach to a two-band model and obtained that $\alpha^{(2)}(\omega_1,\omega_2)\propto(\omega_1+\omega_2)^3/\omega_1^3\omega_2^4$. 
For the case of a $k$-independent dipole matrix element, i.e., assuming that the interband matrix element of the position operator does not depend on $k$,
the dependence obtained here is, however, in agreement with
the result of Ref.~\onlinecite{Aversa1994}, which is based on limiting the general $\chi^{(3)}$ analysis of Ref.~\onlinecite{Aversa1995} to a two-band model.

The band structure enters $\alpha^{(2)}$ explicitly only via the group velocity in field direction.
In the parabolic region near the band gap, Eq.~\eqref{eq:parabolic}, the group velocity is given by
\begin{equation}\label{eq:vk1D}
|v_{k_\mathrm{r}}| = \sqrt{2\hbar(\omega_{k_\mathrm{r}}-\omega_\mathrm{g})/m^*}. 
\end{equation}

The results of Eq.~(\ref{eq:2PA_1D}) can be easily generalized to three dimensions in the parabolic region of the band structure,
since the two directions perpendicular to the field can be included by a continuum of energetically shifted one-dimensional parabolic bands.
These correspond to a constant two-dimensional density of states in the directions perpendicular to the linear polarization of the incident fields.
In this case the factor $|v_{k_\mathrm{r}}|$ in Eq.~(\ref{eq:2PA_1D}) is replaced by
\begin{equation}\label{eq:bandgap-scaling}
(2/3)\sqrt{2\hbar/m^*}(\omega_1+\omega_2-\omega_\mathrm{g})^{3/2},
\end{equation}
while all other factors remain unchanged.
The proportionality of 2PA coefficients to $(\omega_1+\omega_2-\omega_\mathrm{g})^{3/2}$ is well-established from the scaling rules by Wherrett.\cite{Wherrett:84}
The dependence of $\alpha^{(2)}$ on the band gap of the three-dimensional model
is shown in Figure~\ref{fig:2PA_ogap} and again compared with our numerical results.
We find a very good agreement apart from a very small tail below the band gap which simply arises from the finite pulse durations. 

\begin{figure}[t]
	\includegraphics[width=.49\textwidth]{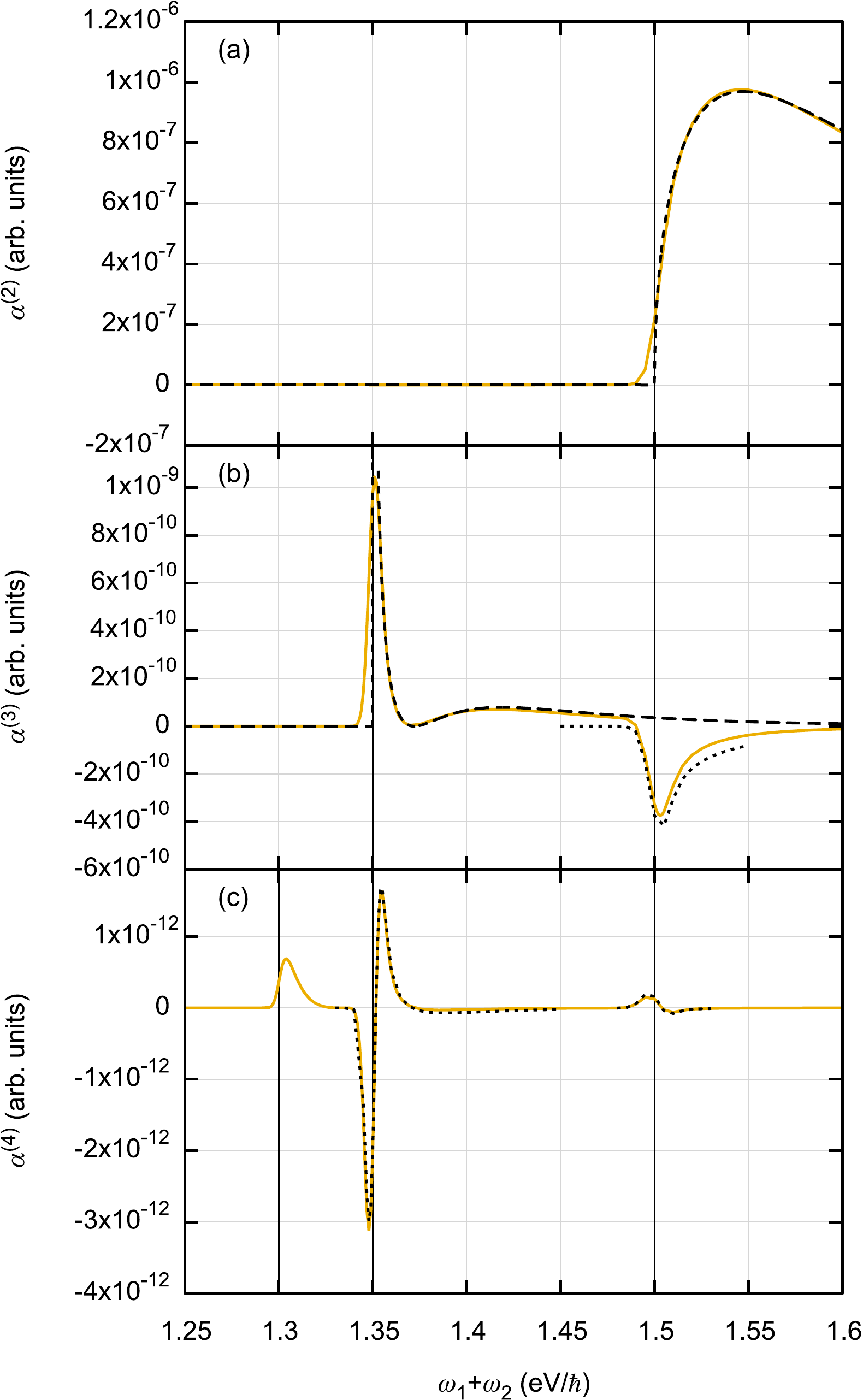}
	\caption{(color online) MPA spectra for $l=2,3,4$ corresponding to the orders $m=3,5,7$ for a one-dimensional tight-binding model \eqref{eq:TBmodel} with $\omega_\mathrm{g}=\omega_\mathrm{b}=$1.5~eV/$\hbar$.
	The probe frequency is fixed at $\omega_1=1.2$~eV$/\hbar$ and the pump frequency is varied.
	The pulse durations for solid lines are $\tau_1=\tau_2=200$~fs and both dephasing and relaxation are switched off. The dashed lines are analytical results, Eq.~\eqref{eq:2PA_1D} in (a) and Eq.~\eqref{eq:3PA_1D} in (b).
	The dotted lines are approximations for the higher-order correction features according to Eq.~\eqref{eq:Starkshift} with $G\approx$ const.
	The vertical thin dashed lines indicate MPA resonances with the band gap, i.e., $\omega_1+(l-1)\omega_2=\omega_\mathrm{g}$. }
	\label{fig:MPA}
\end{figure}

The band-filling factor $(1-n_{\mathbf{k}}^\mathrm{(0)})$ appearing in $\alpha^{(2)}$ is the same as for the 1PA coefficient and implies
a corresponding change from 2PA to stimulated emission when population inversion is present. 
Also for this case we find good agreement between our analytical and numerical results as is shown in Figure~\ref{fig:N0--2PA}
for a three-dimensional two-band GaAs model with initial quasi-equilibrium Fermi-Dirac carrier distributions corresponding to a carrier density of $N_{eh}$,
which is equal for electrons and holes, with a temperature of $T=300$~K.
Due to the strongly different effective masses of the conduction and the heavy-hole valence bands,
$\alpha^{(2)}$ drops steeply to 0 due to the filling of the conduction band near the $\Gamma$ point until $N_{eh} \lesssim 5 \cdot 10^{18}$/cm$^3$.
Then $\alpha^{(2)}$ slowly becomes negative and the 2PA coefficient approaches the negative of the value without initial carriers present.
The plot also highlights the cancellation of the terms depending on the $k$-gradient of the carrier density in the two susceptibility paths $iie$ and $iei$ as well as in the two conductivity paths $eei$ and $eie$.

In general, we see that the dynamic response for quite short pulses is in the considered frequency range and resolution
nearly identical to the continuous wave response, apart from minor deviations arising from the finite spectral width of the pulses. 

\subsection{Higher-order response and multi-photon absorption}
\label{sec:MPA}

We now discuss the probe absorption in the presence of more intense pump pulses where higher-order processes are relevant.
The absorption is again calculated using the perturbative equations (\ref{eq:SBEexp1st})-(\ref{eq:SBEexpansion_n})
including also an expansion in the relevant directions.
This procedure is carried out both numerically for short pulses and analytically for continuous-wave excitation.
Figure~\ref{fig:MPA} shows spectra separately for the three orders $m=3$, $5$, and $7$, as function of the pump frequency $\omega_2$ for a fixed probe frequency of $\hbar \omega_1=1.2$~eV for the one-dimensional tight-binding model given in Eq.~\eqref{eq:TBmodel}.
We use the sum frequency $\omega_1 + \omega_2$ as a common axis and indicate MPA resonances with the band gap by vertical thin dashed lines. 

In third order, see Figure~\ref{fig:MPA}(a), the TPA vanishes below the band gap except the tail arising from the finite spectral width of the pulses.
Above the gap the numerically calculated spectrum agrees well with the analytical result of Eq.~\eqref{eq:2PA_1D} shown by the dashed line (scaled).
After an initial increase which is determined by the group velocity $|v_{k_\mathrm{r}}|$ in the parabolic regime, Eq.~\eqref{eq:vk1D},
the absorption decreases with further increasing sum frequency $\omega_1 + \omega_2$.
This decrease arises from the decrease of the ratio $\omega_1 / \omega_2$, i.e., we move towards 
a more degenerate situation for which according to Fig.~\ref{fig:2PA_freq} the TPA is minimal.

In fifth order, see Figure~\ref{fig:MPA}(b), 3PA starts at $\omega_1 + \omega_2=1.35$~eV$/\hbar$, i.e., when $\hbar \omega_1 + 2 \hbar \omega_2$ is equal to the band gap.
Significant 3PA is present mainly in two separate frequency ranges and nearly vanishes in between (here for $\omega_1+2\omega_2 \approx 1.54$~eV$/\hbar$).
For the used value of the dipole matrix element $\mu = 3 e\,$\AA~the paths with the lowest power of $\mu$ are dominant. According to Figure~\ref{fig:path-tree} three paths are quadratic in $\mu$ when the sample is initially unexcited, while all other paths are either of higher order in $\mu$ or are irrelevant. 
Solving the perturbative expansion for the three dominant paths analytically, we find the following contribution to the 3PA coefficient in the continuous-wave limit for an initially unexcited one-dimensional system
\begin{align}
\label{eq:3PA_1D}
\alpha^{(3)}(\omega_1,\omega_2) &\propto \mu^2
\big[512\, \omega _1^3 \omega _2^8 \left(\omega _1+\omega _2\right)^2 \omega _{k_\mathrm{r}}'\big]^{-1}
 \nonumber\\
& \quad   \cdot
\big[ 
 \omega _1 \omega _2 \left(\omega _1^2+\omega _1 \omega _2+4 \omega _2^2\right) \omega _{k_\mathrm{r}}''
 \nonumber\\
& \quad \ \; 
-2 \left(\omega _1+\omega _2\right) \left(\omega _1^2+2 \omega _2^2\right) \omega _{k_\mathrm{r}}'{}^2
\big]{}^2 .
\end{align}
In the derivation of these expressions we assumed that the sum frequency $\omega_1+\omega_2$ is off-resonant and therefore the fifth-order corrections to 2PA are not included in this result. The frequency tail of the solid line at $\omega _1+2 \omega _2 \lesssim \omega_\mathrm{g}$ which represent the numerical results for pulsed excitation is not included either, since it is again due to the finite spectral width of the pulses. 
Otherwise there is again very good agreement with the numerical result. 
Indeed, the analytical solution \eqref{eq:3PA_1D} has a zero of order 6 (for a parabolic band) in the relevant frequency range. In the degenerate case $\omega_1=\omega_2$, the zero is found at
$\omega_1=2\omega_k'{}^2/ \omega_k''$.

The difference between the numerical and the analytical results at $\omega_1+\omega_2 \gtrsim \omega_\mathrm{g}$ is due to the abovementioned fifth-order corrections to 2PA. Even though its shape is not dispersive it results, when added to the TPA, in a reduction and an optical Stark shift of the 2PA near the band gap to higher frequencies with increasing pump intensity. 
In the next order $m=7$, see Figure~\ref{fig:MPA}(c), further corrections occur to both 2PA and 3PA.
For the chosen frequency values, the Stark-shift correction of the 3PA peak at $\omega_1+2\omega_2\gtrsim \omega_\mathrm{g}$ to higher frequencies is dominant, while there is no significant correction to the broad 3PA range above. 
We would like to point out that the shape of the higher-order corrections to 2PA and 3PA roughly correspond
to a derivative of the lower-order absorption coefficient with respect to the probe frequency. In other words, the correction can be written as
\begin{equation}\label{eq:Starkshift}
	\alpha^{(l+1)}(\omega_1,\omega_2) = G(\omega_1,\omega_2) \frac{\partial}{\partial \omega_1} \alpha^{(l)}(\omega_1,\omega_2),
\end{equation}
where $G(\omega_1,\omega_2)$ varies slowly over the frequency range of the feature, presumably since it contains nonresonant factors only.
Here we do not discuss analytical Stark shift expressions to confirm this relation for a crystal. 
However, consider a two-level system with transition energy $\omega_0$ excited by two beams where $\omega_1$ is near-resonant and $\omega_2$ is detuned.
Then
\begin{equation}\label{eq:Starkshift_TLS}
\alpha^{(l)}(\omega_1,\omega_2) = - [ 2^{2l-1} (\omega_1+\omega_2+i\gamma_n)^{l-1}  (\omega_1-\omega_0+i\gamma_p)^l ]^{-1}
\end{equation}
and \eqref{eq:Starkshift} holds with $G(\omega_1,\omega_2)=-[ 4l (\omega_1+\omega_2+i\gamma_n)]^{-1}$ (the term $\propto\partial G/\partial \omega_1$ is negligible). 
As long as the features are confined to a narrow frequency range the derivative alone fits our numerical results surprisingly well.
The corresponding curves, scaled separately for each feature, are shown by the dotted lines in Figure~\ref{fig:MPA}(b)-(c). 

For resonant excitation the absorption changes as measured in pump-probe experiments typically alternate with the order, e.g., as  demonstrated in Ref.~\onlinecite{Meier2000a} the fifth-order contribution is proportional to the negative of the third-order result.
Interestingly, the relation  given in \eqref{eq:Starkshift}, i.e., that the next higher order is proportional to the derivative of the lower order qualitatively
also holds for the results shown in Ref.~\onlinecite{Meier2000a} when exciting with co-circularly polarized pulses below the
exciton resonance since the third-order Stark shift becomes a broadening in fifth order.

In Figure~\ref{fig:MPA}(c) we also recognize the genuine 4PA with a maximum slightly above $\omega_1+3\omega_2=\omega_\mathrm{g}$, i.e.,
$\omega_1+\omega_2 =$1.3~eV, 
 which is restricted to a quite narrow frequency range.
Finally, we would like to note that further calculations have shown that the MPA coefficients have a similar strong enhancement for strongly non-degenerate excitation as the 2PA coefficient shown in Figure~\ref{fig:2PA_freq}. However, the exact dependence on the frequency ratio obviously differs, in particular, the minima occur for $\omega_1<\omega_2$.

\begin{figure}[t!]
	\includegraphics[width=.49\textwidth]{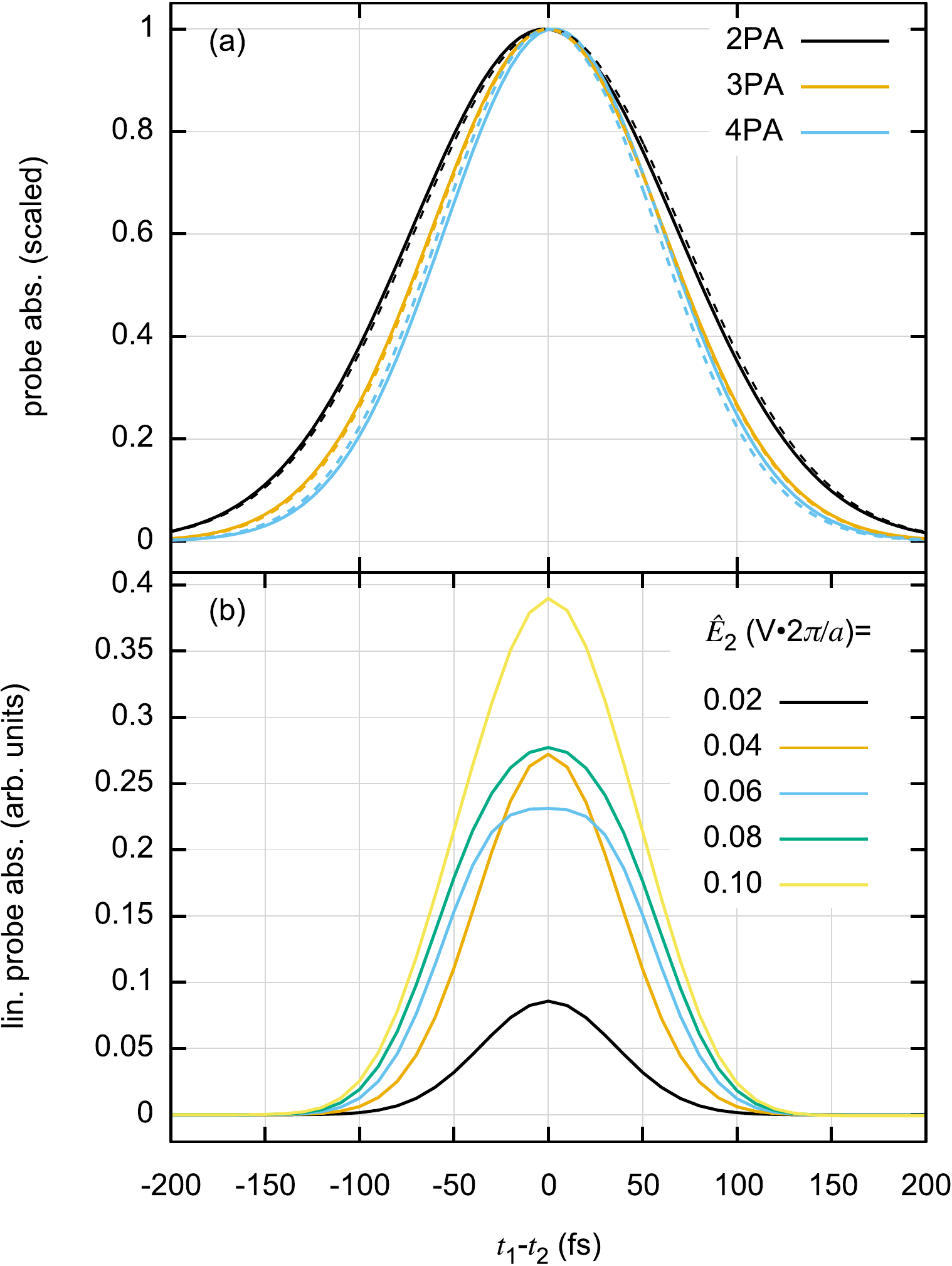}
	\caption{(color online) Transients obtained for the one-dimensional tight-binding model \eqref{eq:TBmodel} with $\omega_\mathrm{g}=\omega_\mathrm{b}=$1.5~eV/$\hbar$. (a): Absorption, $\int \mathrm{d}t\, \mathbf{j}^{(2l-1)} \cdot \mathbf{E}_1$, 
	of the probe pulse for $l=2,3,4$ as a function of the time delay between the probe and pump pulse.
	The parameters are $\hbar\omega_1=1.2$~eV, $\hbar\omega_2=0.31$~eV, $\tau_1=\tau_2=50$~fs, and $T_2\to\infty$.
	The dashed lines correspond to Eq.~(\ref{eq:delay}).
	(b): Transients obtained for strong pump fields (peak amplitudes indicated). $\hbar\omega_1=1.1$~eV, $\hbar\omega_2=0.5$~eV, $\tau_1=\tau_2=50$~fs, and $T_2=200$~fs.
	}
	\label{fig:delay--PrAbs}
\end{figure}

Let us now discuss the influence of a time delay between the pump and the probe pulse on the MPA.
In bulk GaAs 2PA transients have been measured in Ref.~\onlinecite{Penzkofer1989}.
Evaluating the denominator of Eq.~(\ref{eq:PrAbs_pulse}) for two Gaussian pulses, one finds that the $l$PA depends on the time delay $t_1-t_2$ as
\begin{align}
\label{eq:delay}
\exp\left(-\frac{2(l-1)(t_1-t_2)^2}{(l-1)\tau_1^2+\tau_2^2}\right).
\end{align}
Note that  for $l\gg 1$ the temporal width is mainly determined by the duration of the probe pulse. 
Figure~\ref{fig:delay--PrAbs}(a) shows that this scaling is well fulfilled for $\tau_1=\tau_2=50$~fs for the three cases $l=2,3,4$.

Another influence to be discussed for MPA is that of an initial carrier density $n_{q\mathbf{k}}^\mathrm{(0)}$.
Using quasi-equilibrium distributions we find that the entire MPA spectra of Figure~\ref{fig:MPA} scale very well with the band filling factor $(1-n_{\mathbf{k}}^\mathrm{(0)})$.
It is thus excepted that in the continuous wave limit of MPA (using $\gamma_p\to 0$ and $\gamma_n\to 0$) this factor describes the
entire density dependence as for 1PA and 2PA, at least as long as a perturbative treatment is valid.

\begin{figure}[t!]
	\includegraphics[width=.49\textwidth]{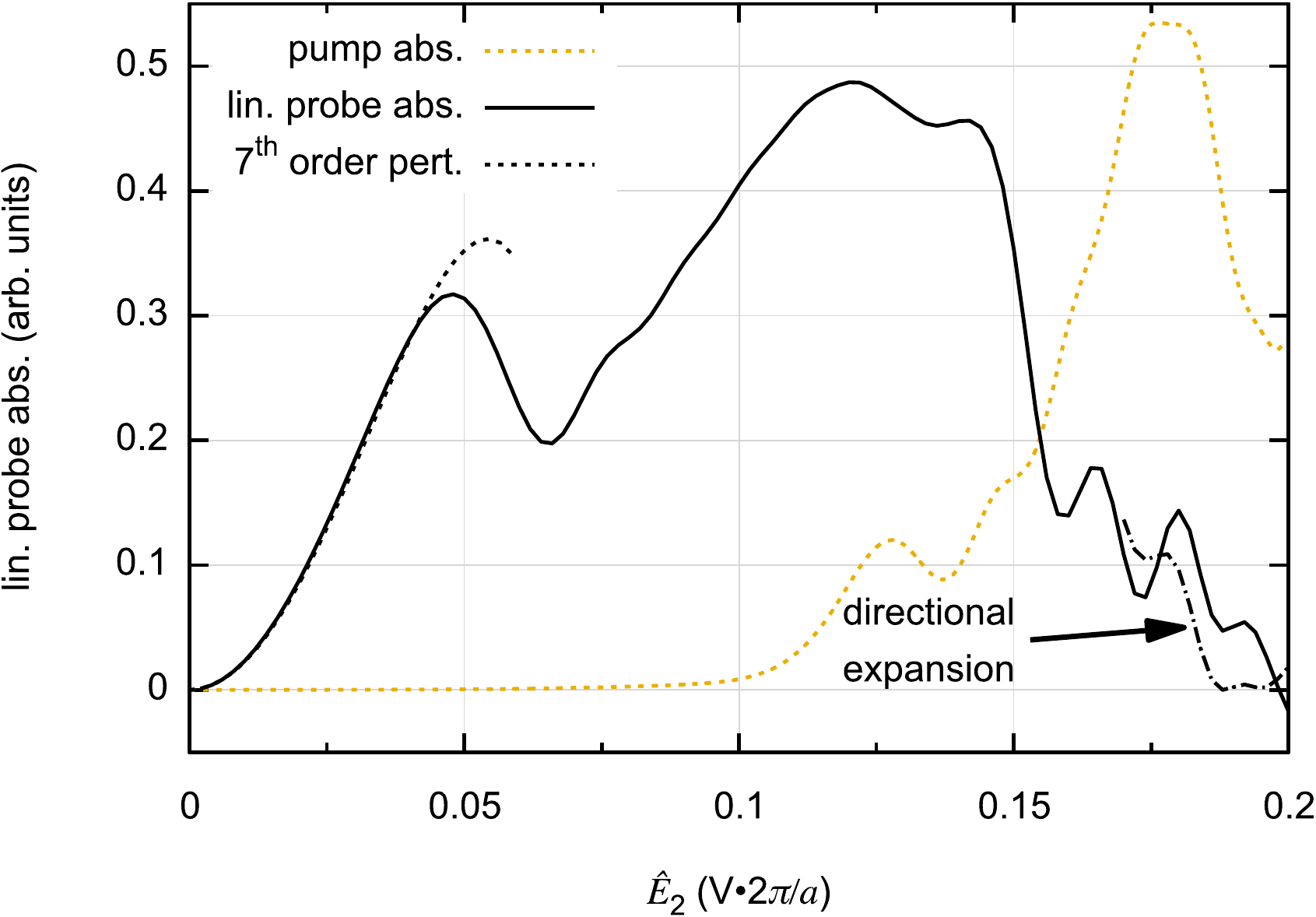}
	\caption{(color online) Absorption of the weak probe pulse as a function of the pump field amplitude ($\hbar\omega_1=1.1$~eV, $\hbar\omega_2=0.5$~eV, $\tau_1=\tau_2=50$~fs, $T_2\to\infty$). The black dotted line is the result from seventh order perturbation theory. The yellow dotted line is the total pump absorption (on a different scale). The one-dimensional tight-binding model~\eqref{eq:TBmodel} with $\omega_\mathrm{g}=\omega_\mathrm{b}=$1.5~eV/$\hbar$ was used.
	}
	\label{fig:E2--PrAbs}
\end{figure}

\subsection{Non-perturbative regime with high pump intensities}
\label{sec:intensepump}

In this section we present results for the probe absorption obtained from the set of SBE (\ref{eq:mixSBE_first})-(\ref{eq:mixSBE_last})
which is applicable for arbitrarily strong pump fields. 
In this regime, the linear probe absorption shows a complex dependence on both the time delay $t_1-t_2$, see Fig.~\ref{fig:delay--PrAbs}(b), and the pump field amplitude $\hat{E}_2$, see Fig.~\ref{fig:E2--PrAbs}).
As we include the probe field to first order, the absorption of probe photons by the simultaneous absorption of one (or more) pump photons does not cause saturation effects itself. The irregular characteristics reflects the population dynamics induced by the strong pump pulse, here mainly in the region of the Brillouin zone that is resonant with $\omega_1+\omega_2$. 
Therefore, the pump absorption is shown on a different scale in Figure~\ref{fig:E2--PrAbs} (analogous to Figure~\ref{fig:1col_abs}). Roughly-speaking, the increase of the pump absorption starting at $\hat{E}_2\approx 0.15$V$\cdot 2\pi/a$ results in a decrease of the probe absorption. More precisely, the dips in the probe absorption are caused by high pump-induced carrier densities $n_{\mathbf{k}}(t)$, where $\mathbf{k}$ is in the $\omega_1+(l-1)\omega_2$-resonant region of the Brillouin zone (where $l \geq 2$) and $t$ is near the peak time $t_1$. 
For instance, the curve for $\hat{E}_2= 0.06$V$\cdot 2\pi/a$ in Figure~\ref{fig:delay--PrAbs} is flattened compared to a Gaussian, and correspondingly a dip is visible near this $\hat{E}_2$-value in Figure~\ref{fig:E2--PrAbs}.

The data in Fig.~\ref{fig:E2--PrAbs} is again obtained for the 1D tight-binding model, \eqref{eq:TBmodel},  since the entire Brillouin zone becomes important for strong fields. 
For weak fields, the absorption completely agrees with the perturbative regime (the seventh order expansion is accurate up to roughly $\hat{E}_2\approx 0.04$V$\cdot 2\pi/a$).
For very strong pump intensities (near the right edge of Figure~\ref{fig:E2--PrAbs}), Eqs.~(\ref{eq:mixSBE_first})-(\ref{eq:mixSBE_last}) are not accurate anymore, since signals in directions different from the probe direction are mixed into the calculated absorption. 
The proper directional expansion is shown in a small region of the plot in Figure~\ref{fig:E2--PrAbs} only since it is computationally expensive.
Note that the probe absorption may take positive or negative values in the regime of very intense pump fields. 
Overall, the non-monotonous intensity dependence is related to Rabi oscillations which occur in regions of the Brillouin zone which are resonantly excited by MPA and the negative regions can be understood to arise from (transient) population inversion.
In fact, the origin and the non-monotonous behavior and possible sign changes of the absorption are similar to effects that have been predicted for
two-color photocurrents.\cite{Pasenow2008}

\section{Summary}
\label{sec:summary}

Using a semiclassical approach based on the semiconductor Bloch equations, we have studied nonlinear optical absorption from one or two light beams within a two-band model. 
In general, the presence of far off-resonant light waves makes it necessary to consider both inter- and intraband excitation effects. 
In the length gauge of the electromagnetic field both types take a simple form which facilitates perturbative analytical solutions of the Bloch equations. 
The imaginary (real) part of the obtained susceptibility (conductivity) together determine the MPA coefficient for the steady state response. 
As long as the order of the perturbation expansion is sufficiently high, the MPA coefficients are in good agreement with absorption spectra obtained numerically for pulsed excitation. 

Our approach, however, is not limited to the perturbative regime. We also demonstrate how to obtain the absorption of a single strong pulse or a weak probe pulse in the presence of an intense pump pulse in a non-perturbative fashion and demonstrate, e.g., a strongly non-monotonous behavior of the intensity-dependent optical absorption which also appears in the transient response.

The simplicity of the employed model enables us to analyze and point out several novel fundamental aspects.
Future extensions of this model might lead to a more realistic description. 
The evaluation of nonlinear optical susceptibilities in more sophisticated band structure models including more than two bands and the $k$-dependence
as well as the vector character of the matrix elements would lead to more quantitative predictions and would also allow to study
the complete tensor character of the nonlinear susceptibilities.
Furthermore, the inclusion of electron-electron-interactions is expected to lead to an excitonic enhancement of MPA appearing spectrally below resonances originating from the fundamental band gap.

\acknowledgments
We thank John E. Sipe for several useful discussions.
Support by the Deutsche Forschungsgemeinschaft (DFG, German Research Foundation) – project number 231447078 – TRR 142 (project A07)
is gratefully acknowledged.
We also thank the PC$^2$ (Paderborn Center for Parallel Computing) for providing computing time.

\bibliography{../bib/bib}

\appendix

\section{Path evolution equations}
\label{sec:paths}

Splitting the evolution equations (\ref{eq:SBEexp1st})-(\ref{eq:SBEexpansion_n}) into individual paths of interband and intraband terms, the series takes the following form in the first two orders:
\begin{align}
\frac{\partial}{\partial t} {p}_\mathbf{k}^{(1)[e]} &= \frac{i}{\hbar}\mathbf{d}\cdot \mathbf{E}\, \big(1-n_{\mathbf{k}}^\mathrm{(0)} \big) - i ({\omega}_\mathbf{k}-i\gamma_p)\, p_\mathbf{k}^{(1)[e]} \\
\frac{\partial}{\partial t} {n}_{q\mathbf{k}}^{(1)[i]} &= \frac{e}{\hbar}\mathbf{E}\cdot \nabla_\mathbf{k}\, n_{q\mathbf{k}}^\mathrm{(0)} - \gamma_n n_{q\mathbf{k}}^{(1)[i]}\\
\frac{\partial}{\partial t} {p}_\mathbf{k}^{(2)[ie]} &= \frac{e}{\hbar}\mathbf{E}\cdot \nabla_\mathbf{k}\, p_\mathbf{k}^{(1)[e]} - i ({\omega}_\mathbf{k}-i\gamma_p) p_\mathbf{k}^{(2)[ie]} \\
\frac{\partial}{\partial t} {p}_\mathbf{k}^{(2)[ei]} &= 
- \frac{i}{\hbar}\mathbf{d}\cdot \mathbf{E}\, n_{\mathbf{k}}^\mathrm{(1)[i]}  - i ({\omega}_\mathbf{k}-i\gamma_p) p_\mathbf{k}^{(2)[ei]} \\
\frac{\partial}{\partial t} {n}_{q\mathbf{k}}^{(2)[ee]} &= 
\frac{i}{\hbar}\mathbf{d}\cdot \mathbf{E}\big(p_\mathbf{k}^{*(1)[e]}\!-p_\mathbf{k}^{(1)[e]}\big) 
- \gamma_n n_{q\mathbf{k}}^{(2)[ee]} \\
\frac{\partial}{\partial t} {n}_{q\mathbf{k}}^{(2)[ii]} &= 
 \frac{e}{\hbar} \mathbf{E}\cdot \nabla_\mathbf{k} n_{q\mathbf{k}}^{(1)[i]} 
- \gamma_n n_{q\mathbf{k}}^{(2)[ii]}
\end{align}
where the path labels are given in square brackets.
The higher orders can be obtained analogously.

\section{Third-order response functions} 
\label{app:phi3}

Here we give explicit expressions for the functions $\phi_\mathbf{k}^{(\mathrm{path};\ell)}(\omega_1,\omega_2)$ defining the third-order susceptibility
and conductivity in the continuous-wave limit according to Eq.~(\ref{chi3_pp}).
We define primed quantities  as
derivatives with respect to $k$ along the field direction, e.g.,
$\omega'_\mathbf{k} = \hat{\mathbf{e}}_1\cdot \nabla_\mathbf{k}\, \omega_\mathbf{k}$,
and similarly for $n_\mathbf{k}$.
The subpath index $\ell$ runs from 1 to 3 and we define
$\omega_3\equiv\omega_2$, $\Omega_1=\Omega_2\equiv \omega_1+\omega_2$, and $\Omega_3\equiv 0$.
Note that the occurrences of these frequency expressions are not invariable, but change according to the arguments of $\phi_\mathbf{k}^{(\mathrm{path};\ell)}(\omega_1,\omega_2)$.
For simplicity we neglect all terms which do not feature a resonance denominator 
$\omega_1-\bar{\omega}_\mathbf{k}$ or $\omega_1+\omega_2-\bar{\omega}_\mathbf{k}$.
This approximation is sufficient for considering 2PA, electronic Raman effect (where $\omega_2 \to -\omega_2$ according to $\eta=-$ in Eq.~(\ref{chi3_pp})), and ac Stark shift. 

\begin{widetext} 
For the four susceptibility paths we find: 
\begin{align}
\phi_\mathbf{k}^{({iie};\ell)}(\omega_1,\omega_2) &= \frac{e^2\mu^2}{8\hbar^3}\,\frac{1}{\omega_1-\bar{\omega}_\mathbf{k}}\,\frac{1}{\Omega_\ell-\bar{\omega}_\mathbf{k}}
\left(
\frac{2\big(1-n_\mathbf{k}^{(0)}\big){\omega'_\mathbf{k}}^2}{(\omega_\ell-\bar{\omega}_\mathbf{k})^{3}} 
+
\frac{\big(1-n_\mathbf{k}^{(0)}\big)\omega''_\mathbf{k}-2{n'_\mathbf{k}}^{\!(0)}\omega'_\mathbf{k}}{(\omega_\ell-\bar{\omega}_\mathbf{k})^{2}} 
-
\frac{{n''_\mathbf{k}}^{(0)}}{\omega_\ell-\bar{\omega}_\mathbf{k}} 
\right)\nonumber
\\
&+
\frac{e^2\mu^2}{8\hbar^3}\,\frac{1}{\omega_1-\bar{\omega}_\mathbf{k}}\, \frac{1}{(\Omega_\ell-\bar{\omega}_\mathbf{k})^2}
\left(
\frac{\big(1-n_\mathbf{k}^{(0)}\big){\omega'_\mathbf{k}}^2}{(\omega_\ell-\bar{\omega}_\mathbf{k})^{2}} 
-
\frac{{n'_\mathbf{k}}^{\!(0)}\omega'_\mathbf{k}}{\omega_\ell-\bar{\omega}_\mathbf{k}}  
\right)  \\
\phi_\mathbf{k}^{({eee};\ell)}(\omega_1,\omega_2) &= 
-\frac{\mu^4}{2\hbar^3}\,\frac{1}{\omega_1-\bar{\omega}_\mathbf{k}}\,
\frac{(\omega_\ell+i\gamma_p)\big(1-n_\mathbf{k}^{(0)}\big)}{(\Omega_\ell+i\gamma_n)(\omega_\ell-\bar{\omega}_\mathbf{k})(\omega_\ell+\bar{\omega}_\mathbf{k}^*)}\\
\phi_\mathbf{k}^{({iei};\ell)}(\omega_1,\omega_2) &= 
-\frac{e^2\mu^2}{8\hbar^3} \,\frac{1}{\omega_1-\bar{\omega}_\mathbf{k}}\, \frac{1}{\omega_\ell+i\gamma_n} \left(\frac{{n''_\mathbf{k}}^{(0)}}{\Omega_\ell-\bar{\omega}_\mathbf{k}}
+ \frac{{n'_\mathbf{k}}^{\!(0)}\omega'_\mathbf{k}}{(\Omega_\ell-\bar{\omega}_\mathbf{k})^2}\right)   \\
\phi_\mathbf{k}^{({eii};\ell)}(\omega_1,\omega_2) &= 
-\frac{e^2\mu^2}{8\hbar^3} \,\frac{1}{\omega_1-\bar{\omega}_\mathbf{k}}\,
\frac{{n''_\mathbf{k}}^{(0)}}
{(\Omega_\ell+i\gamma_n)(\omega_\ell+i\gamma_n)}
\end{align}
The solutions for the four conductivity paths read:
\begin{align}
\phi_\mathbf{k}^{({eie};\ell)}(\omega_1,\omega_2) &= 
\frac{e^2\mu^2}{8\hbar^3} \, \frac{i\omega'_\mathbf{k}}{(\omega_1+i\gamma_n)(\Omega_\ell-\bar{\omega}_\mathbf{k})}
\left(
\frac{\big(1-n_\mathbf{k}^{(0)}\big){\omega'_\mathbf{k}}}{(\omega_\ell-\bar{\omega}_\mathbf{k})^{2}}
-
\frac{{n'_\mathbf{k}}^{\!(0)}}{\omega_\ell-\bar{\omega}_\mathbf{k}}
\right)
 \\
%
\phi_\mathbf{k}^{({iee};\ell)}(\omega_1,\omega_2) &= 
-\frac{e^2\mu^2}{4\hbar^3} \, 
\frac{i\omega'_\mathbf{k}}{(\omega_1+i\gamma_n)(\Omega_\ell+i\gamma_n)}
 \,
\left(
\frac{\big(1-n_\mathbf{k}^{(0)}\big){\omega'_\mathbf{k}}}{(\omega_\ell-\bar{\omega}_\mathbf{k})^2}
-
\frac{{n'_\mathbf{k}}^{\!(0)}}{\omega_\ell-\bar{\omega}_\mathbf{k}}
\right)
\\
%
\phi_\mathbf{k}^{({eei};\ell)}(\omega_1,\omega_2) &= 
-\frac{e^2\mu^2}{8\hbar^3} \, \frac{i\omega'_\mathbf{k}}{(\omega_1+i\gamma_n)(\Omega_\ell-\bar{\omega}_\mathbf{k})}\,
\frac{{n'_\mathbf{k}}^{\!(0)}}{(\omega_\ell+i\gamma_n)} 
\\
%
\phi_\mathbf{k}^{({iii};\ell)}(\omega_1,\omega_2) &= 0 
\end{align}
\end{widetext} 

The paths contributing to 2PA (or Raman for $\eta=-$) are easily identified by the occurrences of the resonance denominator $\omega_1+\omega_2-\bar{\omega}_\mathbf{k}$. 
All paths lacking this term do not contribute to $\alpha^{(2)}$ (despite unphysical divergences for $\gamma_n\to 0$). 
The contributing paths are further evaluated by carrying out the $k$-integral from Eq.~(\ref{chi3_pp}) and taking $\gamma_n\to 0$ and $\gamma_p\to 0$. 
This limit may be immediately carried out in all cases except at the resonance poles. 
According to Eq.~(\ref{eq:2PA}), $\alpha^{(2)}$ is determined by the imaginary part of the poles,
\begin{align}
\mathrm{Im}[(\Omega_1-\bar{\omega}_k)^{-1}] &= \frac{-\gamma_p}{(\Omega_1-\omega_k)^2+\gamma_p^2} \\
\mathrm{Im}[(\Omega_1-\bar{\omega}_k)^{-2}] &= \frac{-2\gamma_p(\Omega_1-\omega_k)}{[(\Omega_1-\omega_k)^2+\gamma_p^2]^2}
\end{align}

As discussed in the main text, we restrict the further evaluation to the one-dimensional case.
We  transform the $k$-integral in Eq.~(\ref{chi3_pp}) into frequency space:
\[ \lim_{\gamma_p\to 0} \int\!\!\mathrm{d}k \ldots = 
\lim_{\gamma_p\to 0} \sum_{k\approx k_\mathrm{r}} \int\!\!\mathrm{d}\omega_k |\omega'_{k}|^{-1} \ldots , \]
where the sum is over the resonant parts of the spectrum.
Then we use the asymptotic expressions
\begin{align}
\lim_{\gamma\to 0}\int\!\!\mathrm{d}x \frac{\gamma}{(x-x_0)^2+\gamma^2} f(x) &= \pi f(x_0) \\
\lim_{\gamma\to 0}\int\!\!\mathrm{d}x \frac{2\gamma(x-x_0)}{[(x-x_0)^2+\gamma^2]^2} f(x) &= \pi f'(x)\big|_{x=x_0} ,
\end{align}
which leads to the result given in Eq.~(\ref{eq:2PA_1D}).

\end{document}